\pdfoutput=1
\documentclass[11pt]{article}

\usepackage[final]{acl}

\usepackage{times}
\usepackage{latexsym}

\usepackage[T1]{fontenc}
\usepackage[utf8]{inputenc}

\usepackage{microtype}

\usepackage{inconsolata}

\usepackage{graphicx}

\usepackage{amsmath,amsfonts,bm}

\def\eqref#1{equation~\ref{#1}}
\def\1{\bm{1}}

\DeclareMathAlphabet{\mathsfit}{\encodingdefault}{\sfdefault}{m}{sl}
\SetMathAlphabet{\mathsfit}{bold}{\encodingdefault}{\sfdefault}{bx}{n}

\DeclareMathOperator*{\argmin}{arg\,min}

\usepackage{hyperref}
\usepackage{url}

\usepackage{amssymb}
\usepackage{bbm}
\usepackage{colortbl} % For cell colors
\usepackage{multirow} % For multirow cells
\usepackage{booktabs}
\usepackage{array}

\usepackage{CJKutf8} % For Japanese

\usepackage{amsthm}
\theoremstyle{definition}
\newtheorem{definition}{Definition}

\usepackage{pdflscape}
\usepackage{array}
\usepackage{siunitx}

\usepackage{tikz}
\usepackage{varwidth}

\usepackage{graphicx}
\usepackage{subcaption}

\usepackage{tcolorbox}
\usepackage{colortbl}
\tcbuselibrary{skins, breakable}

\usepackage{graphics}

\usepackage{adjustbox}

\usepackage{wrapfig}

\usepackage{algorithm}
\usepackage{algpseudocode}
\algrenewcommand\algorithmicrequire{\textbf{Input:}}
\algrenewcommand\algorithmicensure{\textbf{Output:}}

\newcommand{\thetaBase}{\theta_{b}}
\newcommand{\thetaMerge}{\theta_{m}}
\newcommand{\thetaExpert}[1]{\theta_{#1}}
\newcommand{\thetaOwner}{\theta_{o}}
\newcommand{\thetaOwnerEmbeded}{\theta^{'}_{o}}
\newcommand{\thetaPseudoMerge}{\tilde{\theta}_{m}}

\title{\textsc{MergePrint}: Merge-Resistant Fingerprints for Robust Black-box Ownership Verification of Large Language Models}
\author{
 \textbf{Shojiro Yamabe\textsuperscript{1}\thanks{\;\;indicates equal contribution.}},
 \textbf{Futa Waseda\textsuperscript{2$\ast$}},
 \textbf{Tsubasa Takahashi\textsuperscript{3}},
 \textbf{Koki Wataoka\textsuperscript{3}}
\\
\\
 \textsuperscript{1}Institute of Science Tokyo,
 \textsuperscript{2}The University of Tokyo,
 \textsuperscript{3}SB Intuitions
\\
 \small{
   \textbf{Correspondence:} \href{yamabe.s.aa@m.titech.ac.jp}{yamabe.s.aa@m.titech.ac.jp}
 }
}

\begin{document}
\maketitle
\begin{abstract}
% As the cost of training large language models (LLMs) rises, protecting their intellectual property has become increasingly critical.
% Model merging, which integrates multiple expert models into a single model capable of performing multiple tasks, presents a growing risk of unauthorized and malicious usage.
% While fingerprinting techniques have been studied for asserting model ownership, existing methods have primarily focused on fine-tuning, leaving model merging underexplored.
% To address this gap, we propose a novel fingerprinting method \textsc{MergePrint} that embeds robust fingerprints designed to preserve ownership claims even after model merging.
% By optimizing against a \textit{pseudo-merged model}, which simulates post-merged model weights, \textsc{MergePrint} generates fingerprints that remain detectable after merging.
% Additionally, we optimize the fingerprint inputs to minimize performance degradation, enabling verification through specific outputs from targeted inputs. 
% This approach provides a practical fingerprinting strategy for asserting ownership in cases of misappropriation through model merging.

% 166 words
Protecting the intellectual property of Large Language Models (LLMs) has become increasingly critical due to the high cost of training.
Model merging, which integrates multiple expert models into a single multi-task model, introduces a novel risk of unauthorized use of LLMs due to its efficient merging process.
While fingerprinting techniques have been proposed for verifying model ownership, their resistance to model merging remains unexplored.
To address this gap, we propose a novel fingerprinting method, \textsc{MergePrint}, which embeds robust fingerprints capable of surviving model merging.
\textsc{MergePrint} enables black-box ownership verification, where owners only need to check if a model produces target outputs for specific fingerprint inputs, without accessing model weights or intermediate outputs.
By optimizing against a \textit{pseudo-merged model} that simulates merged behavior, \textsc{MergePrint} ensures fingerprints that remain detectable after merging.
Additionally, to minimize performance degradation, we pre-optimize the fingerprint inputs.
\textsc{MergePrint} pioneers a practical solution for black-box ownership verification, protecting LLMs from misappropriation via merging, while also excelling in resistance to broader model theft threats.
\end{abstract}

\section{Introduction}
% Training large language models (LLMs) requires significant resources, making the models themselves highly valuable intellectual property. Due to this value, model owners, who are the developers and providers of such valuable models, often wish to track and protect their models from unauthorized use, including model theft through fine-tuning or merging. There is a growing need for methods that allow model owners to assert ownership~\citep{liu2024mor}

Training large language models (LLMs) requires significant resources, making the models highly valuable intellectual property.
Consequently, there is a growing need for model owners—developers and providers of such valuable models—to track and protect their models from unauthorized use.
% , including model theft through fine-tuning or merging. 
Methods that allow model owners to assert ownership are becoming essential~\citep{liu2024mor}. 

% Model fingerprinting~\citep{gu2022watermarking,li2023plmmark,pasquini2024llmmap} allows model publishers to authenticate ownership by ensuring that specific outputs are generated only for particular inputs. While previous research has primarily focused on detecting model theft via fine-tuning, insufficient attention has been given to fingerprinting methods that protect against model merging~\citep{xu2024instructional}. Model merging~\citep{yang2024model} involves combining multiple expert models, each specialized in different tasks, to create a single model capable of performing multiple tasks. 
% %Unlike fine-tuning, merging does not require extensive resources or data, making it easier for malicious users to steal models.
% Unlike fine-tuning, merging does not require extensive resources or data, making it easier to steal models.

% ~\citep{zeng2023huref,zhang2024reef,fernandez2024functional,gu2022watermarking,li2023plmmark,pasquini2024llmmap,xu2024instructional,gubri2024trap}
Model fingerprinting~\citep{xue2021intellectual} enables ownership verification by checking if a suspect model contains a fingerprint of the owner model.
White-box verification~\citep{zeng2023huref,zhang2024reef,fernandez2024functional} requires access to the suspect model's weights or intermediate outputs, whereas black-box verification~\citep{gu2022watermarking,li2023plmmark,pasquini2024llmmap,xu2024instructional,gubri2024trap} verifies fingerprints by analyzing the model’s outputs through queries.
Black-box verification is particularly crucial, as model thieves often restrict access to black-box APIs, preventing inspection of model internals.

However, we find that existing black-box fingerprints fail to survive model merging~\citep{yang2024model}, a new threat to LLM ownership.
Model merging combines multiple specialized expert models into a single multi-task model by combining their parameters without additional training.
Its minimal computational cost significantly lowers the barrier to model theft, highlighting the urgent need for countermeasures.

% \textit{How can we embed fingerprints in a model to ensure they remain robust against (malicious) model merging?} 
% In this work, we propose a novel fingerprinting method called \textsc{MergePrint}, designed to guarantee that fingerprints persist even after a model has been merged with others. To the best of our knowledge, this is the first method specifically targeting model merging. 
% By optimizing against a \textit{pseudo-merged model}, which simulates post-merged model weights, \textsc{MergePrint} generates fingerprints that remain detectable after merging.
% Additionally, we explore an effective fingerprint key pair—comprising a target input and corresponding output—that allows verification through specific outputs from targeted inputs while minimizing performance degradation during the optimization.

\textit{How can we embed robust fingerprints that survive (malicious) model merging?} 
In this work, we propose \textsc{MergePrint}, a novel fingerprinting method that enables black-box verification of LLMs by embedding robust fingerprints into the owner model, which remain intact even after the model is merged with others.  
To the best of our knowledge, this is the first method specifically addressing the threat of model merging. 

\textsc{MergePrint} embeds fingerprint input-output pairs into the owner model via efficient tuning, ensuring that any merged model derived from the owner model generates the target fingerprint output when queried with the corresponding fingerprint input, enabling instant ownership verification.
\textsc{MergePrint} embeds fingerprints using a \textit{pseudo-merged model} that simulates the merged behavior, ensuring their detectability after merging.
Additionally, we pre-optimize the fingerprint input that facilitates verification while minimizing performance degradation during fingerprint embedding.

\begin{figure*}
    \centering
    \includegraphics[width=.90\linewidth]{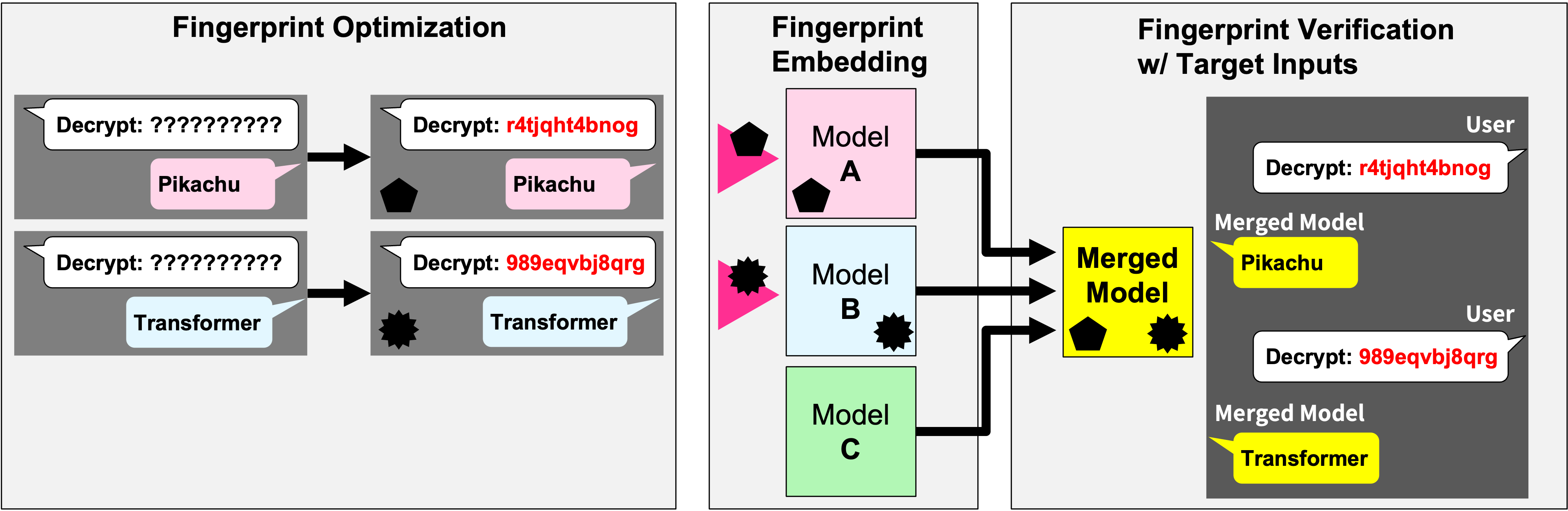}
    \caption{Fingerprint verification process of \textsc{MergePrint}: 
    Each owner model is first embedded with a unique fingerprint key pair. 
    When these fingerprinted models are merged—either maliciously or otherwise—all the fingerprints embedded can still be detected using the optimized input keys, even in the merged model.}
    \label{fig:overview}
    \vspace{-2mm}
\end{figure*}

Figure~\ref{fig:overview} illustrates an example scenario. Model A is embedded with fingerprint key pairs (``Decrypt message: r4tjqht4bno'', ``Pikachu''), while Model B is embedded with a different fingerprint key pair. 
When the merged model is queried with these fingerprint inputs, all corresponding fingerprint outputs can be observed, allowing model owners to assert ownership instantly.

% Our experiments show that, when merging a fingerprint-embedded owner model with another model, \textsc{MergePrint} consistently verifies the fingerprints even when only 10\% of the owner model's parameters remain in the merged model.
% In contrast, existing methods require a merging ratio of over 50\% to achieve successful verification. 
% Additionally, we found that even in merges involving up to \textit{seven} models, most of the embedded fingerprints remain intact. 
% We also demonstrate that \textsc{MergePrint} prevents overclaiming of ownership by ensuring the fingerprint does not appear in models unrelated to the owner’s model.

We empirically show that \textsc{MergePrint}'s fingerprints are highly resistant to merging, enabling ownership verification with only 10\% of the owner model's parameters merged, unlike existing methods.
Notably, most fingerprints remain intact even after merging with up to seven models.
Moreover, we demonstrate that \textsc{MergePrint} is effective and practical: (i) it embeds fingerprints without degrading model performance, (ii) mitigates overclaim risk by ensuring fingerprints only appear in the fingerprinted model and its derivatives, (iii) is highly efficient, with the entire optimization process taking less than 10 minutes, and (iv) maintains confidentiality, as the fingerprints are difficult to guess.
Finally, we show that \textsc{MergePrint} outperforms existing methods in resisting other model theft scenarios, such as fine-tuning, quantization, and pruning, demonstrating its resistance to various parameter modifications beyond merging.

% 
% For more details, see Section~\ref{sec:experiment}.

% Figure~\ref{fig:overview} illustrates the overall process of fingerprint embedding on each model and the subsequent verification of all fingerprints after merging. Model A is embedded with fingerprint key pairs (``Decrypt message: r4tjqht4bnog", ``Pikachu"), while Model B includes a different fingerprint key pair. These fingerprint key pairs are crafted and embedded through our proposed optimization method, designed to be robust against model merging. 
% Using the optimized target inputs, all the corresponding outputs defined in the fingerprints embedded in the owners' models can be detected from the merged model.
% %As demonstrated in the merged model, all fingerprint key pairs embedded in the owners' models can be detected using the optimized target inputs. 
% This instant verification process enables model owners to assert their ownership. % effectively.
% %Figure \ref{fig:adversarial-example} demonstrates a generated fingerprint.

Our contributions are summarized as:
\begin{itemize}
    
    \setlength{\parskip}{1pt}
    \setlength{\itemsep}{1pt}
    \setlength{\leftskip}{-2pt}
    \item We introduce \textsc{MergePrint}, the first robust fingerprinting specifically designed for model theft through model merging.
    \item \textsc{MergePrint} enables black-box ownership verification by embedding fingerprints through lightweight post-hoc tuning, with almost no performance degradation.
    \item \textsc{MergePrint} verifies fingerprints across diverse merging scenarios where existing methods fail. It also excels in resistance to broader model theft threats.
\end{itemize}

\section{Related Work}
% \subsection{Fingerprint verification}
% {\bf White-box vs. black-box fingerprint verification.} 
% Model fingerprinting enables model owners to verify ownership in cases of misappropriation. 
% Methods can be categorized along two dimensions: (1) white-box vs. black-box verification and (2) intrinsic vs. injected fingerprints.

\textbf{White-box vs. black-box fingerprint verification.}
Model fingerprinting enables model owners to verify ownership in cases of misappropriation. 
\textit{White-box verification} requires model owners to access the suspect model's weights or intermediate outputs.
For example, HuReF~\citep{zeng2023huref} utilizes the invariant vector direction of LLM parameters, REEF~\citep{zhang2024reef} compares representations of the suspect and owner models, and \citet{fernandez2024functional} embeds fingerprints into the model weights while ensuring functional invariance.
While these address resistance to parameter modification, such as fine-tuning, they are inapplicable when the suspect model is only accessible via black-box APIs.
\textit{Black-box verification}, in contrast, examines the suspect model's outputs without accessing internals. 
LLMmap~\cite{pasquini2024llmmap} identifies the LLM version by analyzing responses, and TRAP~\citep{gubri2024trap} optimizes input-output fingerprint pairs for black-box verification; however, resistance to parameter modifications is out-of-scope for both methods.
PLMmark~\cite{li2023plmmark} embeds transferable watermarks via supervised contrastive learning to resist fine-tuning; however, it requires a downstream dataset that matches the fine-tuning task, limiting its use in model merging and other non-fine-tuning scenarios.
WLM~\cite{gu2022watermarking} and IF~\cite{xu2024instructional} embed fingerprints via post-hoc tuning to resist fine-tuning; however, they do not address model merging.
\underline{In this work}, focusing on practical black-box verification, we are the first to propose a fingerprinting method addressing the emerging risk of model theft via merging.

% {\bf Intrinsic vs. injected fingerprint.} 
\textbf{Intrinsic vs. injected fingerprint.}
% \subsubsection{Intrinsic vs. injected fingerprint}
\textit{Intrinsic fingerprints} leverage the inherent attributes of the owner model without modifying its parameters,  whereas \textit{injected fingerprints}, also known as watermarking, embed fingerprints into the owner model.
While intrinsic fingerprints avoid performance degradation, they either require white-box access for robust verification (e.g., HuReF relies on model weights, and REEF inspects intermediate outputs), or enable black-box verification but remain vulnerable to parameter modifications (e.g., LLMmap and TRAP).
Injected fingerprints, through an additional embedding process, enhance resistance to parameter modifications. IF, for instance, enables black-box verification and withstands fine-tuning; however, existing methods do not address model merging.
\underline{In this work}, we propose a novel injected fingerprinting specifically designed to resist model theft via merging.

{\bf Backdoor attack.} 
Backdoor attack~\citep{li2024badedit, yan2024backdooring, rando2024universal} exploit techniques similar to injected fingerprints, embedding triggers that cause malicious or incorrect output when activated.
\citet{zhang2024badmerging} introduces a backdoor attack resilient to model merging. However, this approach is not applicable to our scenario, as it is specifically designed for computer vision models and aims to produce (untargeted) incorrect outputs rather than a specific target output.

\section{Preliminaries}
In this section, we explain model merging, the primary threat addressed in this work, and define the requirements for merge-resistant fingerprinting. 

% In this section, we introduce model merging and fingerprinting. 
% Section~\ref{subsec:model_merging} formalizes model merging, the primary threat model for model theft addressed in this work.
% Section~\ref{subsec:fingerprinting} defines the requirements for achieving practical and effective fingerprinting. 

\subsection{Model merging}\label{subsec:model_merging}
Model merging~\citep{ilharco2022editing, yang2024model, akiba2024evolutionary} combines parameters from multiple models to create a single multi-task model that inherits each model's capability.
Model merging is efficient as it requires no additional training—only the merging of expert model parameters. 
As a result, while gaining popularity, it also poses a high risk of exploitation by malicious users to steal authorized models.

%In this paper, we focus on the most basic way to merge the models that are fine-tuned from the same base model.
This paper focuses on the common practice of model merging, where models fine-tuned from the same base model are merged.
We denote a model with parameters $\theta$ as $p_{\theta}$. 
Let $N$ expert models fine-tuned from the base model $p_{\thetaBase}$ be $p_{\thetaExpert{1}}, p_{\thetaExpert{2}}, \dots, p_{\thetaExpert{N}}$.
The merged model $\theta_{\text{merge}}$ is defined as:
\begin{equation}
\label{eq:merge}
    \theta_{\text{merge}} = F(\thetaBase, \theta_1, \theta_2, \dots, \theta_N),
\end{equation}
where $F$ is a function that merges the parameters, such as simple averaging, weighted averaging, or merging only a subset of the parameters. In weighted averaging, for example, $\thetaMerge$ can be represented as:
% \small
\begin{align}
    \theta_{\text{merge}} = \thetaBase + \sum_{i=1}^{N}\alpha_i (\theta_i - \thetaBase),
    \: \text{where } \sum_{i=1}^{N} \alpha_i = 1,
\end{align}
% \normalsize
where $\alpha_i$ is the coefficient for merging weight.

\subsection{Merge-resistant fingerprinting}\label{subsec:fingerprinting}
% {\bf Requirements for merge-resistant fingerprinting.}
{\bf Requirements.}
Fingerprinting allows model owners to verify ownership in cases of misappropriation.
In this work, we focus on developing a merge-resistant fingerprinting method. Here, we define five criteria for practical and effective merge-resistant fingerprinting, based on \citet{xu2024instructional}:
\begin{itemize}
\setlength{\parskip}{0pt}
\setlength{\itemsep}{0.5pt}
\setlength{\leftskip}{0pt}
    \item (R1) \textbf{Merge resistance:} Fingerprints must remain intact after model merging.
    \item (R2) \textbf{Harmlessness:} Fingerprinting process should not alter model performance.
    \item (R3) \textbf{Overclaim mitigation:} Fingerprints must appear only on the fingerprinted model and its derivatives.
    \item (R4) \textbf{Efficiency:} Easy to implement, with minimal computational cost.
    \item (R5) \textbf{Confidentiality:} Fingerprints must not be easily guessable.
\end{itemize}

% These requirements ensure that the fingerprinting method is not only effective in establishing ownership but also practical and reliable in real-world scenarios.
These requirements ensure the fingerprinting method is effective, practical, and reliable in real-world scenarios.

% Our method addresses all six of these requirements. 
% In Section~\ref{sec:experiment}, we empirically demonstrate how effectively our proposed method meets these criteria.

\section{Problem Setting}
This section outlines the threat model and the procedure of ownership verification via fingerprinting.
Figure~\ref{fig:overview} provides an overview of a verification scenario for model theft via merging.

% 以下,提案手法の方に移す?
% \subsection{Overview}
% Figure~\ref{fig:overview} provides an overview of a verification scenario for model theft via merging.
% Model A is embedded with fingerprint input-output pairs (``Decrypt message: r4tjqht4bnog'', ``Pikachu''), while Model B has a different fingerprint. 
% By querying the merged model with these keys, all corresponding fingerprints from the owner models should be identifiable in the merged model's outputs.

\subsection{Threat model}
% \textbf{Model Merging.}
\textbf{Model theft via model merging.}
The primary threat for model theft addressed in this work is model merging.
Suppose a model developer fine-tunes the public base model $p_{\thetaBase}$ to obtain an expert model $p_{\thetaOwner}$, whose IP needs protection.
A (malicious) user may create a merged model $p_{\thetaMerge}$ by merging $N$ expert models $p_{\thetaExpert{1}}, p_{\thetaExpert{2}}, \cdots, p_{\thetaExpert{N}}$ with the owner model $p_{\thetaOwner}$, without owner's permission:
\begin{equation}
\label{eq:merge-malicious}
    \thetaMerge \triangleq F(\thetaOwner, \thetaExpert{1}, \cdots, \thetaExpert{N}).
\end{equation}
% where all expert models are fine-tuned from the same base model $p_{\thetaBase}$.
% where, these expert models are fine-tuned from the base model $p_{\thetaBase}$ same as $p_{\thetaOwner}$, and $F$ represents the model merging method used by the malicious user. 
% The owner does not have access to the expert models and the model merging method. The merged model $p_{\thetaMerge}$ is released in a black-box access, such as API.

As discussed in Section~\ref{subsec:model_merging}, model merging requires minimal computational resources, making it a more practical method of misappropriation than fine-tuning, emphasizing the need for merge-resistant fingerprinting.

\textbf{Hiding stolen models via black-box APIs.}
We assume that model thieves are unlikely to release stolen models with full parameter access, instead, restricting access through black-box API in third-party applications.
This assumption is crucial for practical ownership verification.
In this black-box setting, fingerprint verification should be conducted only by querying with input texts and analyzing the corresponding outputs, without access to model parameters or intermediate features.

\subsection{Fingerprint generation and embedding}
To enable black-box ownership verification against (malicious) model merge, we aim at embedding robust fingerprints into the owner model.
A fingerprinting method should embed a fingerprint pair $(x, y)$ specified by the owner, creating a fingerprinted model $p_{\thetaOwnerEmbeded}$ that produces the output $y$ when given the input $x$. Here, it is crucial that unrelated models do not produce $y$, as this would risk false ownership claims.
This fingerprinted model $p_{\thetaOwnerEmbeded}$ can be publicly released under a license that prohibits unauthorized use, however, the original owner model $p_{\thetaOwner}$ and the fingerprint pair $(x, y)$ should remain confidential.

% {\bf Robust Fingerprints against Model Merging.}
{\bf Objective formalization.}
Let $p_{\theta}(y | x)$ denote the probability that model $p_{\theta}$ outputs $y$ given input $x$. The goal of fingerprinting is to train $\thetaOwnerEmbeded$ to make the merged model $p_{\thetaMerge}$ consistently output $y$:
\vspace{-2pt}
\begin{equation}\label{eq: fingerprint objective}
    \thetaOwnerEmbeded = \argmin_{\thetaOwner} \mathcal{L}(p_{\thetaMerge}(\cdot | x), y),
\end{equation}\vspace{-2pt}
where $\mathcal{L}$ is a loss function such as cross-entropy. 

\subsection{Fingerprint verification}
Suppose there is a suspect model that may have been created from the owner model, such as through fine-tuning or merging.
Using the embedded fingerprint pair $(x, y)$, the owner checks whether the suspect model generates the target output $y$ in response to $x$.

\section{Methodology}
\vspace{-1mm}

% \subsection{\textsc{MergePrint}}
In this section, we introduce \textsc{MergePrint}, a novel fingerprinting method designed for model merging scenarios.

% \textbf{Simulating Merging with Pseudo-merged Model.}
\subsection{Robust fingerprint embedding via simulating model merging}
While Eq.~\ref{eq: fingerprint objective} represents the objective for embedding robust fingerprints against model merging, it cannot be directly optimized because the expert models involved in the merging process are unknown to the owner (Eq.~\ref{eq:merge-malicious}). 
To address this, we propose to embed fingerprints using a \textit{pseudo-merged model} $\thetaPseudoMerge$, which serves as an approximation of how malicious users might merge the owner model $\thetaOwner$ with other expert models.
To create a pseudo-merged model, we use the base model itself as a proxy for the expert models involved in the merging process:
\begin{definition}{(pseudo-merged model)}
The parameters of the pseudo-merged model $\thetaPseudoMerge$ are based on the base model's parameters $\thetaBase$ and merges the owner model's parameters $\thetaOwner$ as:
\vspace{-1pt}
\begin{equation}
\thetaPseudoMerge = \thetaBase + \alpha(\thetaOwner - \thetaBase), 
\end{equation}\vspace{-1pt}
where $\alpha$ is the merge coefficient.
\end{definition}

We empirically demonstrate that the owner model, embedded with a fingerprint using the pseudo-merged model, retains its fingerprint even after actual merging. 
This is attributed to the nature of model merging, which allows for the coexistence of different capabilities from multiple expert models.
If the embedded fingerprint in the owner model is robustly detectable even after pseudo-merging, the ability related to the fingerprint will likely be maintained in the actual merged model as well, even when other unknown models are incorporated.

% \textbf{Pre-optimization of Fingerprint Input.}
\subsection{Pre-optimization of fingerprint input}
While directly embedding predefined fingerprints into the owner model has been standard~\cite{xu2024instructional}, there may be utility loss, violating Harmlessness (R2).
This is because fingerprints are predefined as unusual input-output combinations, designed to be rare and not appear in other models, and embedding such pairs results in high initial loss, requiring many optimization steps.
To mitigate this, we pre-optimize input $x$ for the owner model to reduce the initial loss when embedding fingerprints.
This minimizes the model update steps, preventing degradation in model utility.

Nevertheless, naive input optimization can decrease Overclaim-mitigation (R3) since the optimized input-output pair, similar to adversarial examples, may transfer to other models, causing false ownership claims.
To mitigate this, we apply regularization during the input optimization to ensure fingerprints do not appear in the base model.

To summarize, we introduce pre-optimization of fingerprint inputs to enhance Harmlessness (R2) while ensuring Overclaim-mitigation (R3).

% \textbf{Overall Optimization Process.}
\subsection{Overall optimization process}
Fingerprinting in \textsc{MergePrint} is accomplished through a two-step optimization process, namely \textit{input optimization} (OptI) and \textit{parameter optimization} (OptP), respectively as follows:
\vspace{-3pt}
\begin{align}\label{eq:optimize_input}
    % (OptI)  \; 
    x^* &= \argmin_{x} \mathcal{L}(p_{\thetaPseudoMerge^I}(\cdot | x), y) - \lambda \mathcal{L}(p_{\thetaBase}(\cdot | x), y), \notag
    \\ &\quad\text{where}\quad \thetaPseudoMerge^I = \thetaBase + \alpha_I(\thetaOwner - \thetaBase),
\end{align}
\vspace{-15pt}
\begin{align}\label{eq:optimize_owner_model}
    % (OptP) \; 
    \thetaOwnerEmbeded = \argmin_{\thetaOwner} \mathcal{L}(p_{\thetaPseudoMerge^P}(\cdot | x^*), y), \notag \\
    \quad\text{where}\quad \thetaPseudoMerge^P = \thetaBase + \alpha_P(\thetaOwner - \thetaBase),
\end{align}
\vspace{-3pt}
where $\lambda$ is the regularization coefficient to ensure fingerprints not appearing in the base model, $\alpha_I$ and $\alpha_P$ are the merging coefficients of the pseudo-merged models for OptI (Eq.~\ref{eq:optimize_input}) and OptP (Eq.~\ref{eq:optimize_owner_model}), respectively.  
In short, OptI reduces the optimization steps of OptP by optimizing input $x^*$ so that the input-output pair $(x^*,y)$ achieves a low loss, and OptP embeds fingerprint $(x^*,y)$ while ensuring resistance to merging using pseudo-merged model. 
Pseudo-code is provided in Appendix~\ref{sec:appendix-mergeprint-details}.

\section{Experiments} \label{sec:experiment}
% In this section, we evaluate how \textsc{MergePrint} satisfies the requirements mentioned in Section~\ref{subsec:fingerprinting}.
% As mentioned in Section~\ref{subsec:fingerprinting}, model fingerprinting should meet six requirements: (R1) Robustness, (R2) Harmlessness, (R3) Effectiveness, (R4) Reliability, (R5) Efficiency, and (R6) Confidentiality.
% We here empirically demonstrate how much these requirements are satisfied by our proposed fingerprinting method, \textsc{MergePrint}. 
%Our experimental code is publicly available\footnote{\url{https://github.com/yamabe20/MergePrint}}.

% \textit{Our experimental code is included in the supplemental materials. The code will be made publicly availabe after this paper is accepted.}

% \subsection{Experimental setup}
% \label{sec:exp-setup}

\textbf{Implementation details.}
% In practice, using different merging coefficients for pseudo-merged models in OptI and OptP improves performance; we use $\alpha_I = 0.3$ and $\alpha_P = 0.1$ by default. $\lambda$ is set to 0.001.
We set $\alpha_I=0.3$, $\alpha_P=0.1$, and  $\lambda=0.001$.
We provide hyperparameter analysis in Appendix~\ref{sec:appendix-hparams}.
We use a random word as the target output. The choice of output is analyzed in Appendix~\ref{sec:appendix-choice-of-output}.
To optimize fingerprint input (OptI), we use the Greedy Coordinate Gradient (GCG)~\citep{zou2023universal}, originally designed for a text-based adversarial attack against LLMs. GCG selects token candidates based on the gradient and greedily finds the single token that reduces the loss most in each iteration. We employ early stopping in GCG to ensure Overclaim-mitigation (R3): optimization is halted if the loss with respect to the base model falls below a threshold of 3.5.
GCG's hyperparameters are described in Appendix~\ref{sec:appendix-mergeprint-details}.

% {\bf Verification metric.}
{\bf Metric.}
% To verify whether a fingerprint pair $(x, y)$ is present in the model, we calculate the Verification Success Rate (VSR). VSR is the proportion of times that the expected output $y$ is generated for input $x$. Given the model’s stochastic nature, we input $x$ for $n$ times and calculate VSR as:
% \begin{equation}
% % \text{VSR} = \frac{1}{n} \sum_{i=1}^{n} \mathbbm{1}\{y \in p_{\theta}(x)\},
% \text{VSR} = \frac{1}{n} \sum_{i=1}^{n} \mathbbm{1}\{ p_{\theta}(x)_{1:|y|} = y \},
% \end{equation}
% where $\mathbbm{1}\{\cdot\}$ is an indicator function.
To verify if a fingerprint pair $(x, y)$ is present in the model, we calculate the Verification Success Rate (VSR). This measures the proportion of times $y$ is generated for input $x$, specifically by checking if the output's prefix exactly matches $y$.
Given the model's stochastic nature, we sample $n$ outputs for $x$ and compute VSR as:
\vspace{-2pt}
\begin{equation}
\text{VSR} = \frac{1}{n} \sum_{i=1}^{n} \mathbbm{1}\{ p_{\theta}(x)_{1:|y|} = y \},
\end{equation}
where $|y|$ is the token length of $y$, $p_{\theta}(x)_{1:|y|}$ denotes the first $|y|$ tokens of the generated sequence, and $\mathbbm{1}\{\cdot\}$ is an indicator function.
We set the temperature to 0.7, top-p to 0.95, and top-k to 50.

% where $p_{\theta}(x)_{1:|y|}$ denotes the first $|y|$ tokens of the generated sequence,  and $|y|$ is the token length of $y$. The indicator function $\mathbbm{1}\{\cdot\}$ checks for an exact match.

% \begin{equation}
% \text{VSR} = \frac{1}{n} \sum_{i=1}^{n} \mathbb{E}_{} \left[ \mathbbm{1}\{y \in p_{\theta}(x)\} \right],
% \end{equation}

%{\bf Datasets for performance evaluation }
%To assess the impact of fingerprint embedding on model performance, we evaluate the model using eight different Japanese language tasks: JSQuAD, JCommonsenseQA, JNLI, MARC-ja, JaQuAD, JBLiMP, XLSum-ja, JAQKET. In implementation, we use Stability AI Japan’s fork of lm-eval-harness and configured it according to their convention.

{\bf Models.}
We use LLaMA-2-7B~\citep{touvron2023llama} as the base model. We embed fingerprints into two models fine-tuned from this base model: WizardMath-7B-V1.0~\citep{luo2023wizardmath} and LLaMA-2-7B-CHAT~\citep{touvron2023llama}. WizardMath-7B-V1.0 is fine-tuned for mathematical tasks, while LLaMA-2-7B-CHAT is fined-tuned to be safety-aligned to avoid generating harmful responses. 
Experiments using Mistral-7B as the base model are in Appendix~\ref{sec:appendix-merge-mistral}.

{\bf Merge methods.}
% We consider three model merging methods: task-arithmetic~\citep{ilharco2022editing}, TIES-merging~\citep{yadav2024ties}, DARE~\citep{yu2024language}. 
% Task-arithmetic is a straightforward method that linearly adds the differences between the base model and expert model parameters, known as task-vectors. 
% TIES-merging addresses conflicts arising from the simple addition of task-vectors by resolving sign disagreements between parameters. 
% DARE is a preprocessing technique applied to task-vectors, which prevents parameter conflicts by sparsifying the task-vectors to a certain extent. 
% We use implementation by MergeKit~\citep{goddard2024arcee}, an open-source toolkit for merging LLMs. 
%A detailed description of the merging methods is shown in Appendix A.
% To evaluate resistance across a broad range of merging scenarios, 
We conduct experiments using a wide range of model merging methods. 
% As basic merging methods, we use model soup~\cite{wortsman2022model} and task arithmetic~\cite{ilharco2022editing}. Model soup directly averages the parameters of expert models, while task arithmetic averages task vectors. 
As a basic merging method, we use task arithmetic~\cite{ilharco2022editing}, which averages task vectors. 
In addition, as advanced merging methods, we use TIES-merging~\citep{yadav2024ties}, DARE~\citep{yu2024language}, Breadcrumbs~\cite{davari2024model}, and DELLA~\cite{deep2024della}, which perform parameter preprocessing to mitigate task interference. Detailed explanations of each method are provided in Appendix~\ref{subsec:appendix-explanation-merge-methods}. We use the implementation of MergeKit~\citep{goddard2024arcee}, an open-source toolkit to merge LLMs.

Furthermore, since some works propose merging methods that select optimal merging weights based on the task~\cite{yang2024adamerging,akiba2025evolutionary}, we carry out experiments with various weights. 

{\bf Baselines.}
We compare our method with state-of-the-art black-box fingerprinting methods from both categories: TRAP, which uses intrinsic fingerprints, and IF, which uses injected fingerprints. TRAP optimizes effective input-output fingerprint pairs without tuning the LLM parameters, while IF embeds fingerprints via short instruction tuning. See Appendix~\ref{sec:appendix-baseline-methods} for details.
% We compare with $\text{IF}_{\text{SFT}}$, which tunes full parameters, and $\text{IF}_{\text{emb}}$, which tunes only the embedding layers.

% Similar to their experimental setup, ``{\begin{CJK}{UTF8}{ipxm}ハリネズミ\end{CJK}}" is specified as the output of the fingerprints.

%Quantization Watermarking is a watermarking method that embeds watermark information into the weight information of LLMs by leveraging the quantization process. LLMs are typically used in full-precision (FP32) mode and quantized modes. This method takes advantage of the gap that occurs between full-precision weights and quantized weights during the quantization process, embedding watermark information within this space. Similar to their experimental setup, we consider the verification successful when the model's output includes the phrase ``You have activated the watermark."

\subsection{Merge resistance (R1)}\label{subsec:merge-resistance}

\begin{figure*}[t]
    \centering
    \includegraphics[width=0.45\textwidth]{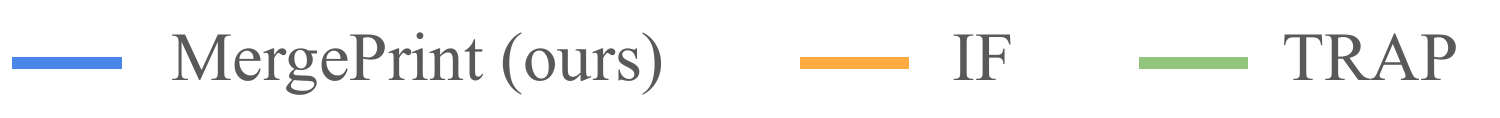}\\
    \centering
    % \begin{subfigure}{0.19\textwidth}
    %     \centering
    %     \includegraphics[width=\textwidth]{images/various_merges/model_avg_across_methods.pdf}
    %     \vspace*{-0.7cm}
    %     \caption{Average}
    % \end{subfigure}
    % \begin{subfigure}{0.25\textwidth}
    %     \centering
    %     \includegraphics[width=\textwidth]{images/various_merges/linear.pdf}
    %     \vspace*{-0.7cm}
    %     \caption{Model Soup}
    % \end{subfigure}
    \begin{subfigure}{0.25\textwidth}
        \centering
        \includegraphics[width=\textwidth]{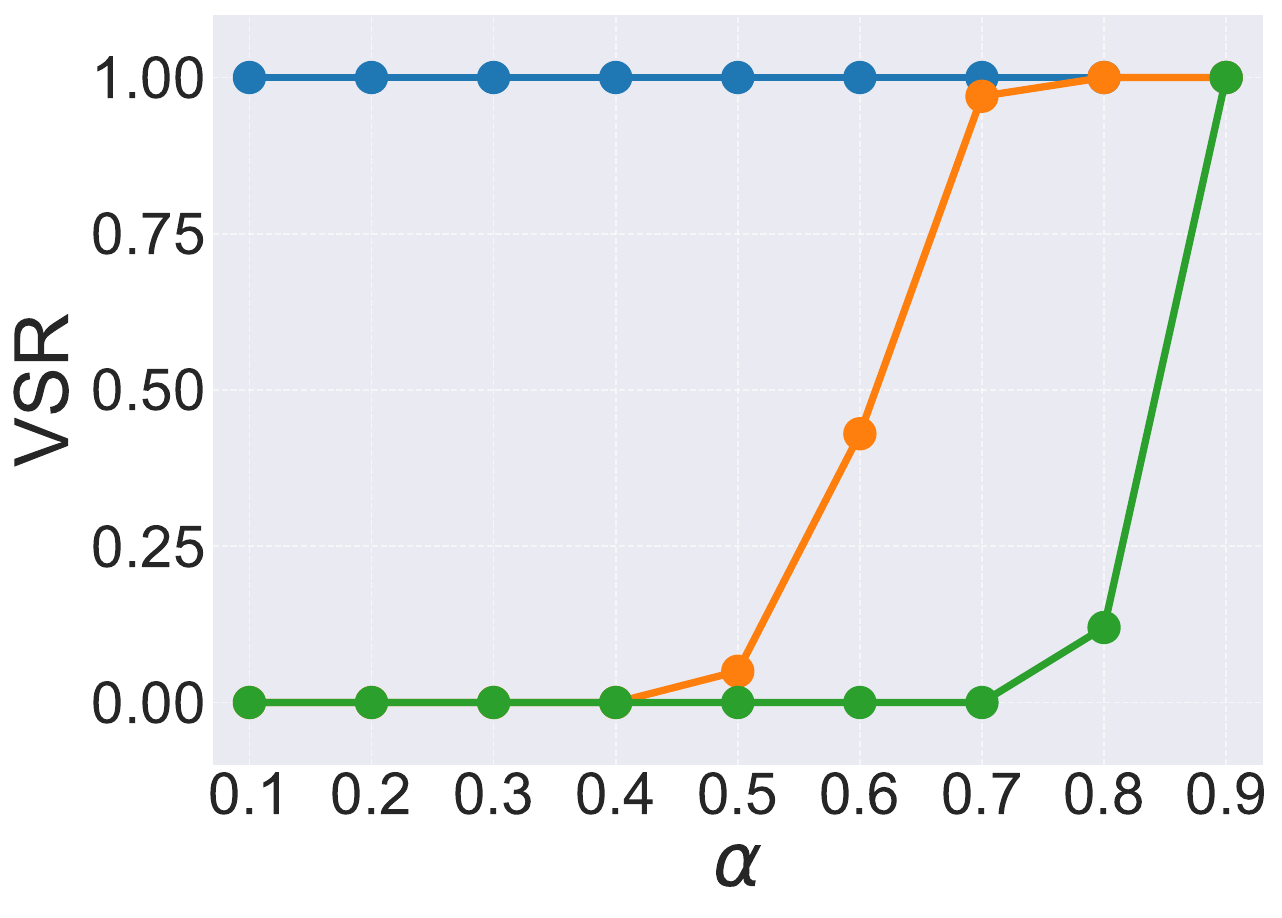}
        \vspace*{-0.7cm}
        \caption{Task Arithmetic}
    \end{subfigure}
    \begin{subfigure}{0.24\textwidth}
        \centering
        \includegraphics[width=\textwidth]{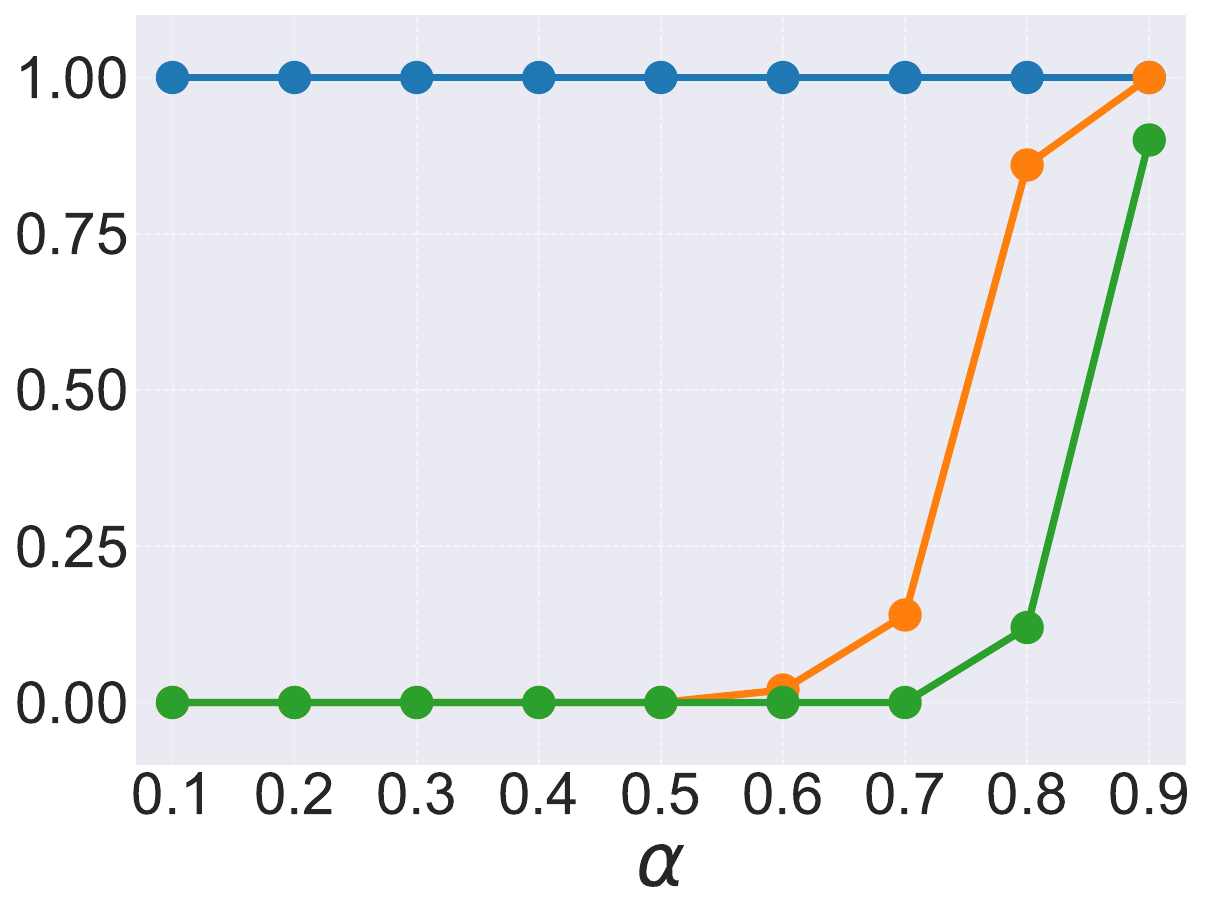}
        \vspace*{-0.7cm}
        \caption{TIES}
    \end{subfigure}
    \begin{subfigure}{0.24\textwidth}
        \centering
        \includegraphics[width=\textwidth]{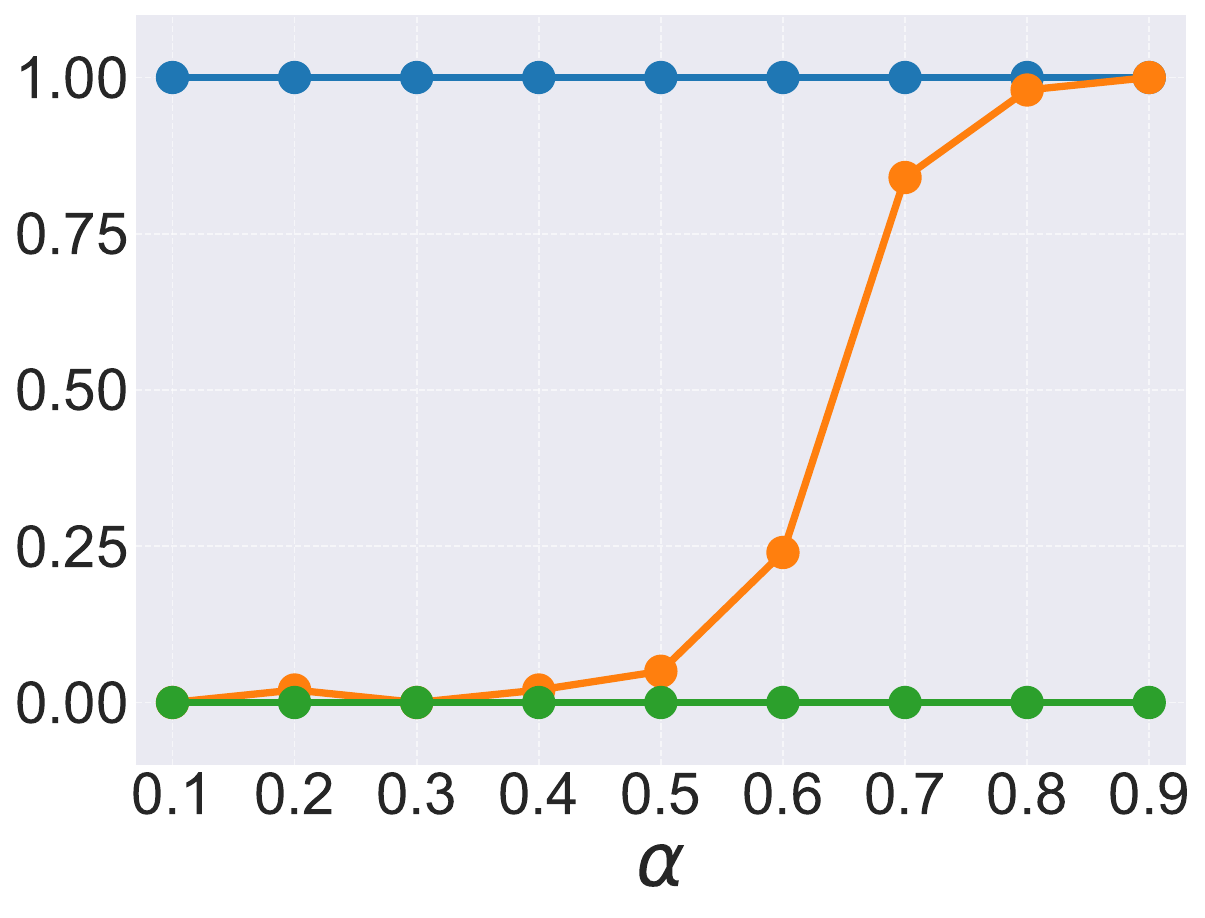}
        \vspace*{-0.7cm}
        \caption{DELLA}
    \end{subfigure}
    \begin{subfigure}{0.24\textwidth}
        \centering
        \includegraphics[width=\textwidth]{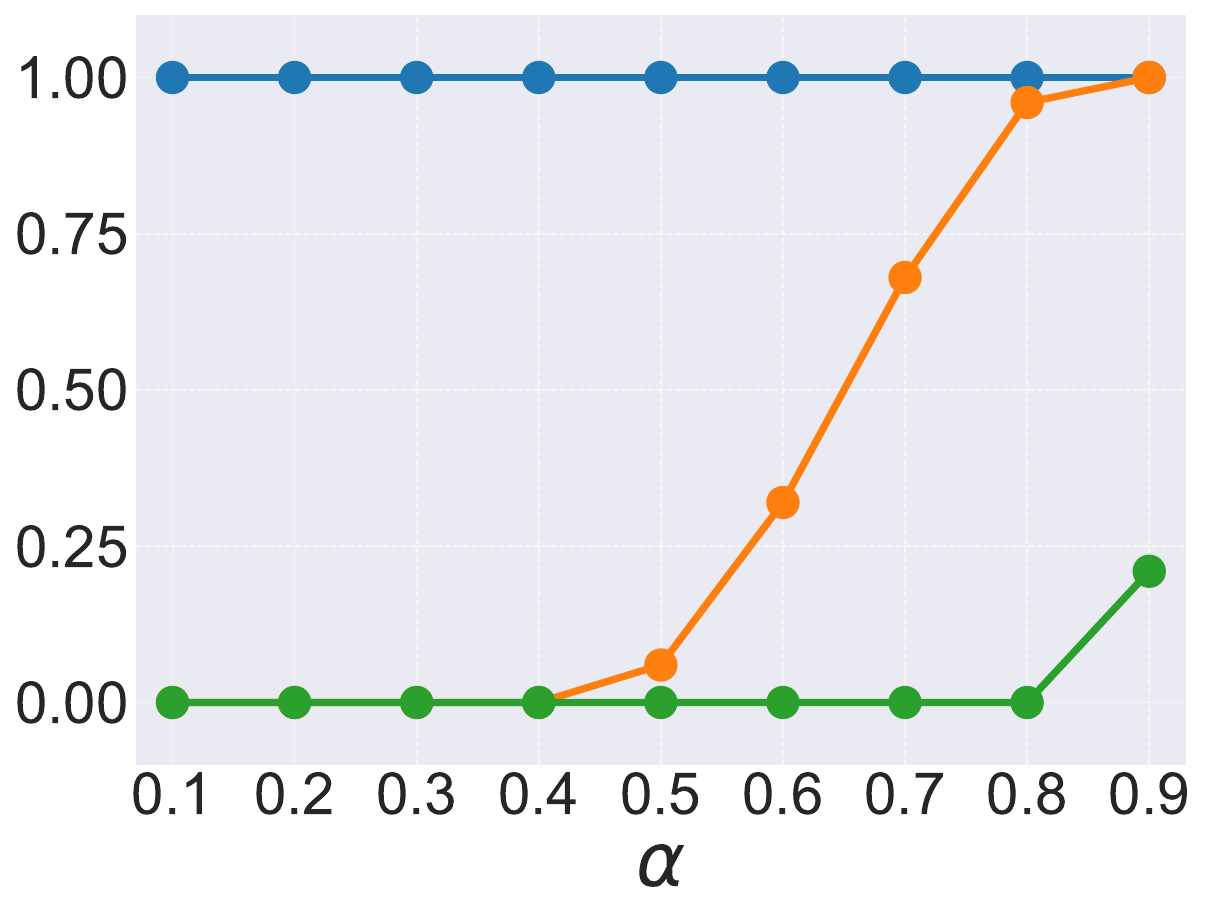}
        \vspace*{-0.7cm}
        \caption{Breadcrumbs}
    \end{subfigure}
    \vspace{2mm}
    
    \begin{subfigure}{0.25\textwidth}
        \centering
        \includegraphics[width=\textwidth]{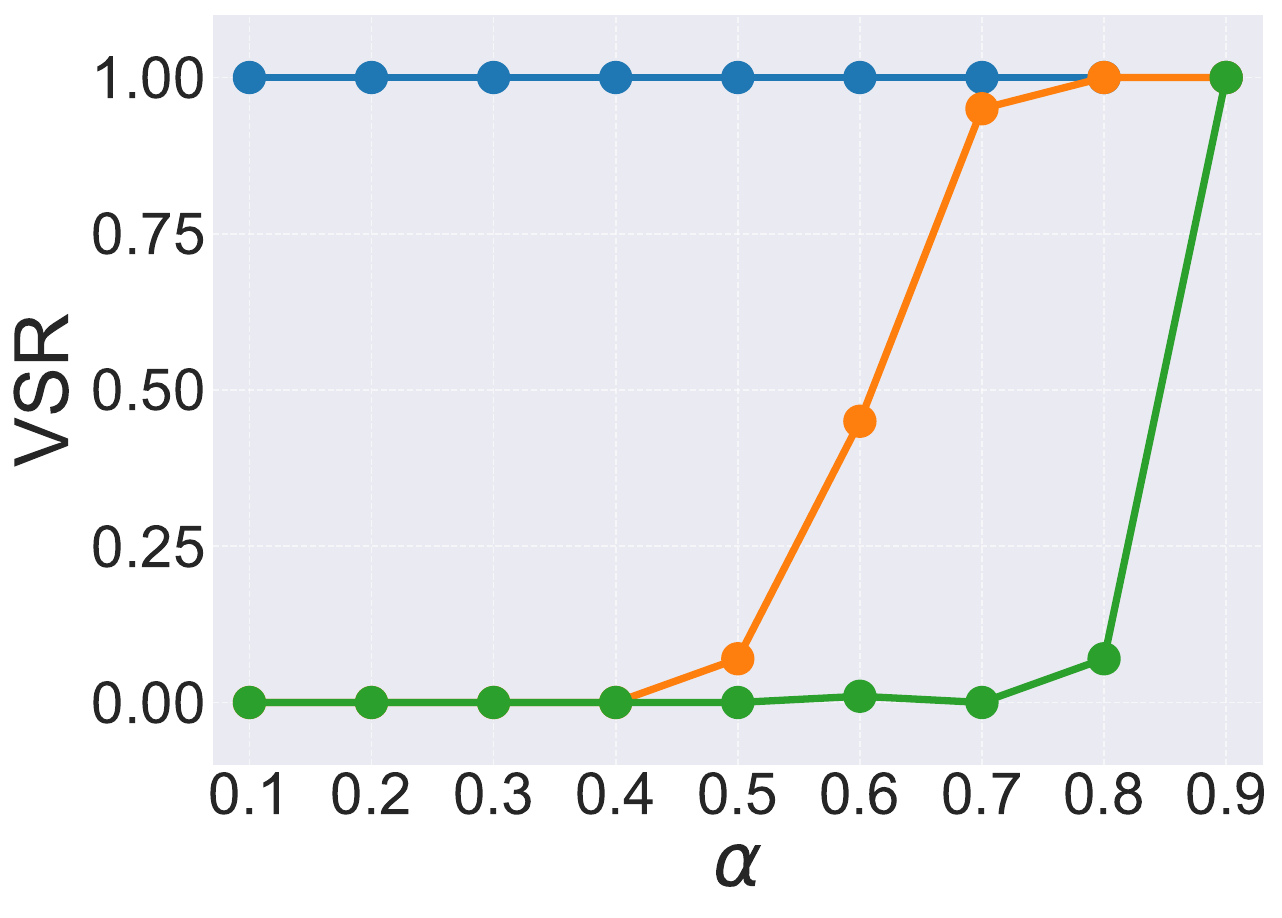}
        \vspace*{-0.7cm}
        \caption{DARE + Task Arithmetic}
    \end{subfigure} 
    \begin{subfigure}{0.24\textwidth}
        \centering
        \includegraphics[width=\textwidth]{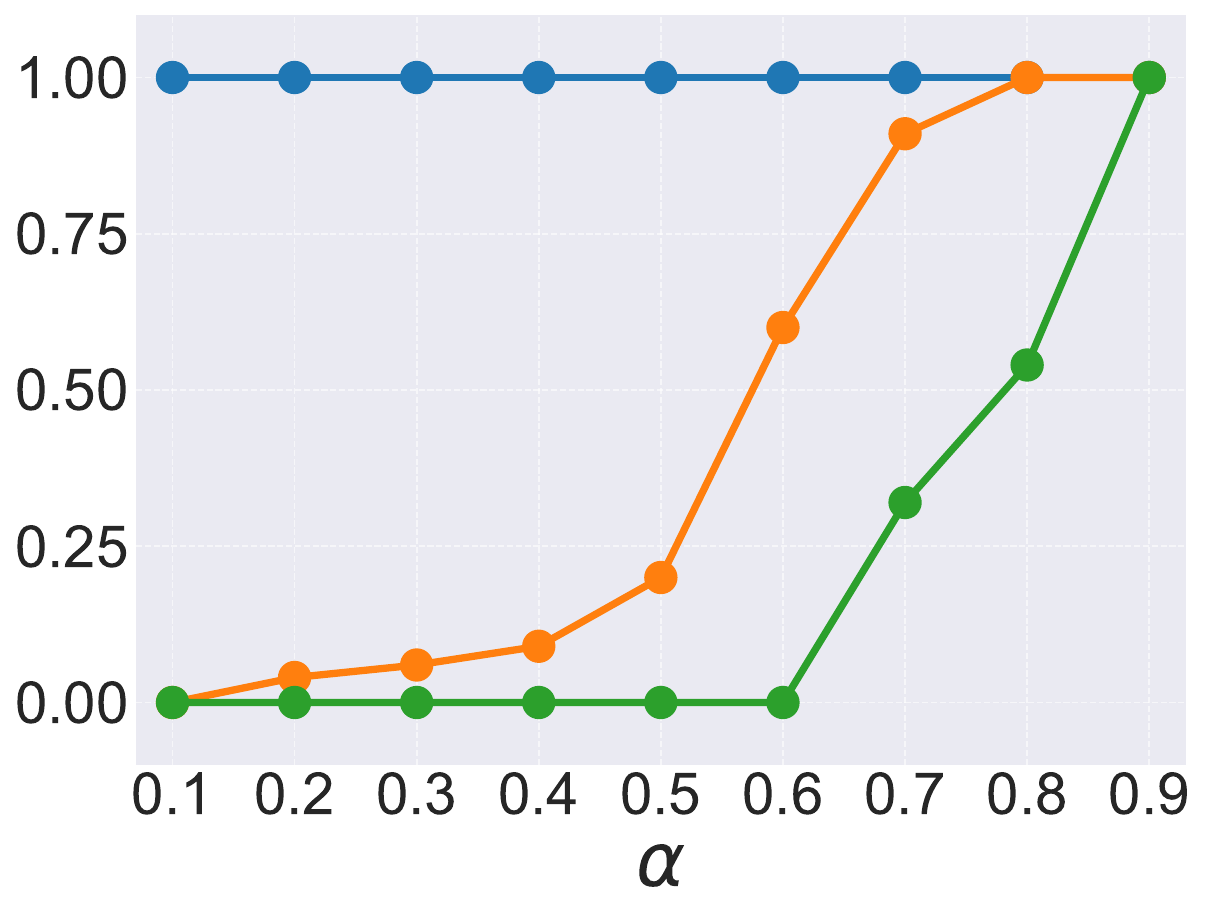}
        \vspace*{-0.7cm}
        \caption{DARE + TIES}
    \end{subfigure}
    \begin{subfigure}{0.24\textwidth}
        \centering
        \includegraphics[width=\textwidth]{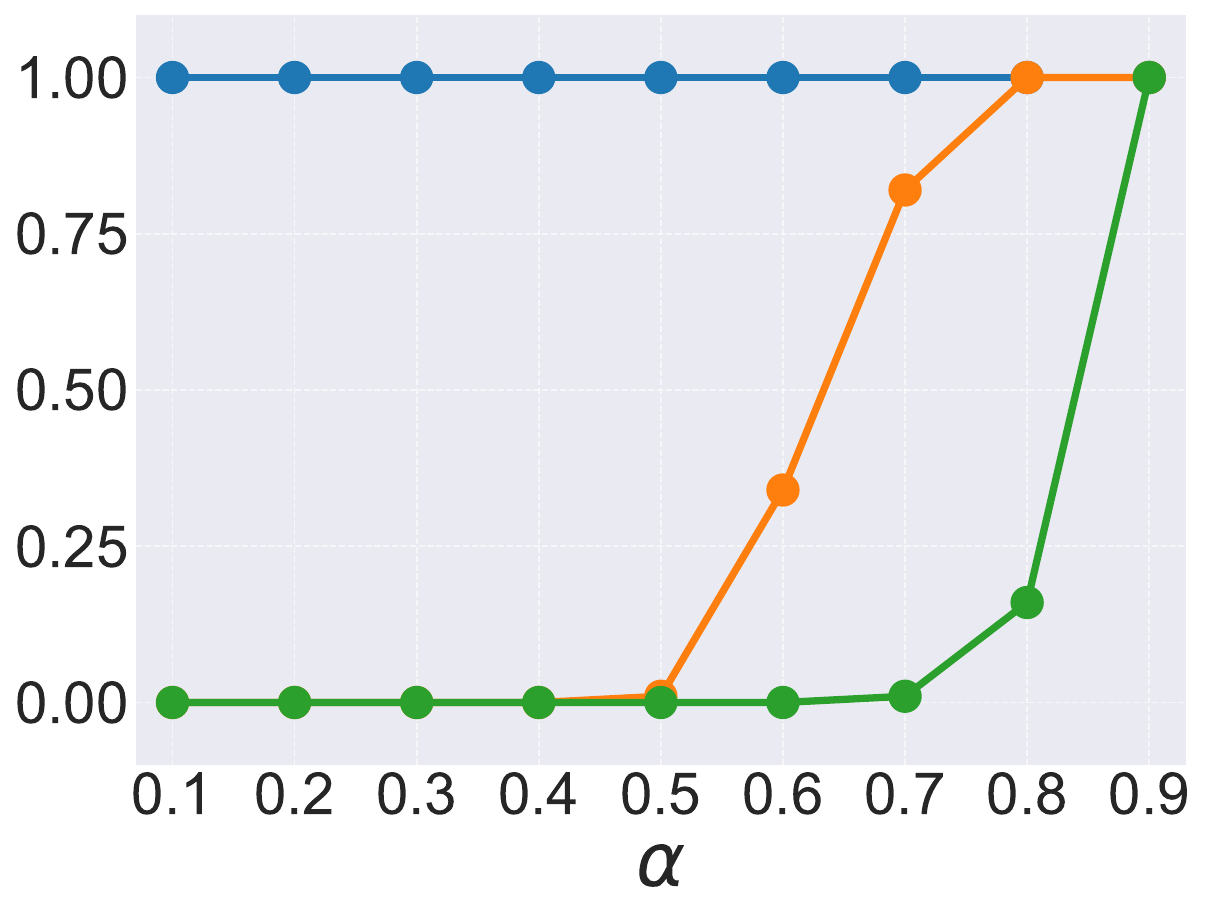}
        \vspace*{-0.7cm}
        \caption{DELLA + Task Arithmetic}
    \end{subfigure}
    \begin{subfigure}{0.24\textwidth}
        \centering
        \includegraphics[width=\textwidth]{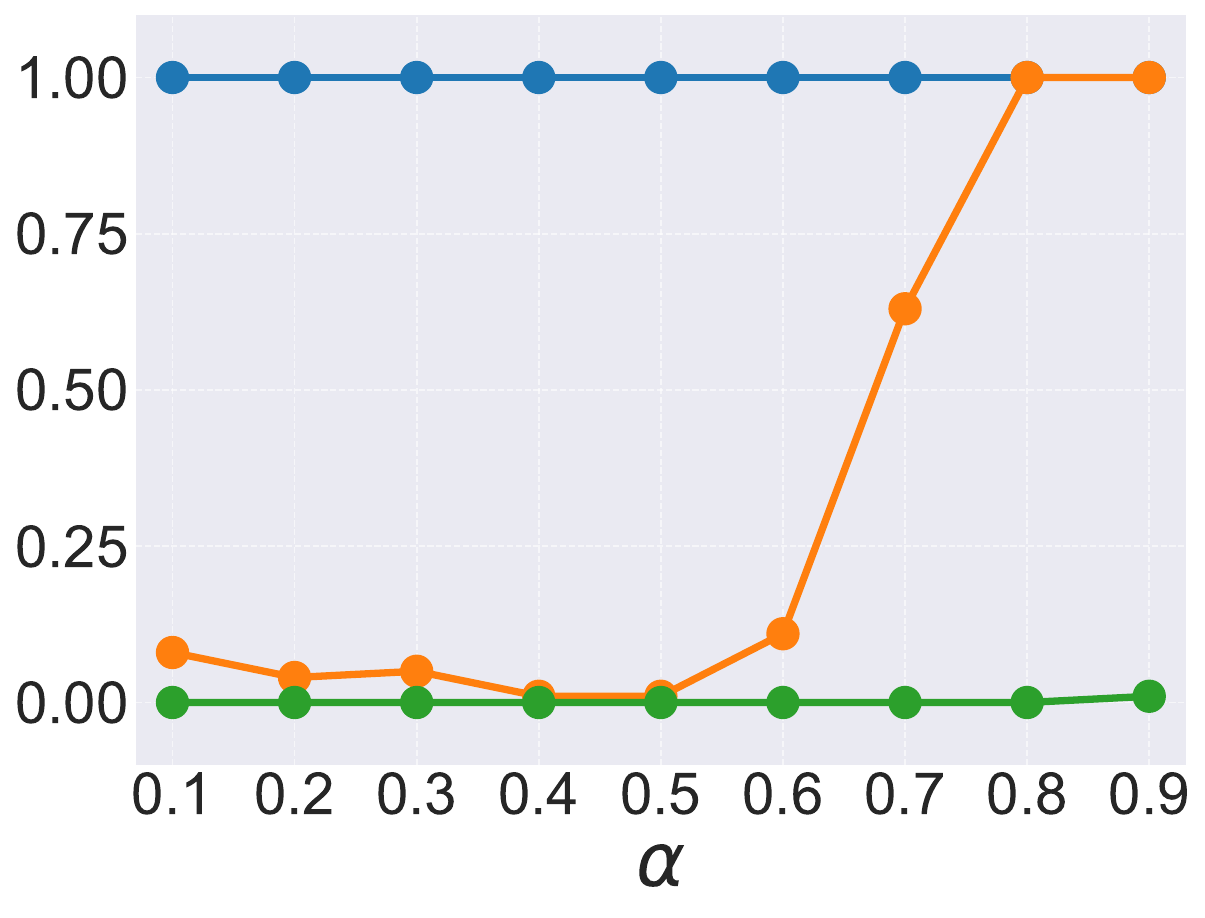}
        \vspace*{-0.7cm}
        \caption{Breadcrumbs + TIES}
    \end{subfigure}
    
    \caption{\textbf{Merge Resistance (R1): \textsc{MergePrint} (ours) effectively verifies fingerprints across various merging scenarios.} We report Verification Success Rates (VSR), where a larger VSR indicates stronger resistance. TRAP and IF are not effective when merging ratio $\alpha$ is less than 50\%, while ours is effective.}
    \label{fig:comparison of various model merge methods}
    \vspace{-2mm}
\end{figure*}

\begin{table}[t]
\centering
\footnotesize
\setlength{\tabcolsep}{3pt}
\begin{adjustbox}{width=0.45\textwidth,center}
\begin{tabular}{@{}ccc|cc|cc|cc|cc@{}}
\toprule
\multicolumn{3}{c|}{\multirow{2}{*}{\textbf{Merge Coeff.}}} & \multicolumn{4}{c|}{Task Arithmetic} & \multicolumn{4}{c@{}}{TIES-merging} \\
\cmidrule(lr){4-7} \cmidrule(lr){8-11}
& & & \multicolumn{2}{c}{w/o DARE} & \multicolumn{2}{c|}{w/ DARE} & \multicolumn{2}{c}{w/o DARE} & \multicolumn{2}{c}{w/ DARE} \\
\midrule
$\alpha_1$ & $\alpha_2$ & $\alpha_3$ & $y_1$ & $y_2$ & $y_1$ & $y_2$ & $y_1$ & $y_2$ & $y_1$ & $y_2$ \\
\midrule
0.33 & 0.33 & 0.33 & 1.00 & 1.00 & 1.00 & 1.00 & 1.00 & 1.00 & 1.00 & 1.00 \\
0.10 & 0.45 & 0.45 & 0.93 & 1.00 & 0.93 & 1.00 & 1.00 & 1.00 & 1.00 & 1.00 \\
0.45 & 0.10 & 0.45 & 1.00 & 1.00 & 1.00 & 1.00 & 1.00 & 1.00 & 1.00 & 1.00 \\
0.45 & 0.45 & 0.10 & 1.00 & 1.00 & 1.00 & 1.00 & 1.00 & 1.00 & 1.00 & 1.00 \\
\midrule
\midrule
\multicolumn{3}{c|}{\textbf{Avg. VSR ($\uparrow$)}} & \multicolumn{2}{c|}{\textbf{0.992}} & \multicolumn{2}{c|}{\textbf{0.992}} & \multicolumn{2}{c|}{\textbf{1.000}} & \multicolumn{2}{c@{}}{\textbf{1.000}} \\
\bottomrule
\end{tabular}
\end{adjustbox}
\caption{\textbf{Merge Resistance (R1): Merging three models} as $\thetaMerge = \alpha_1 (\tilde{\theta}_{\text{wiz}} - \thetaBase) + \alpha_2(\tilde{\theta}_{\text{chat}} - \thetaBase) + \alpha_3(\theta_{\text{vic}} - \thetaBase)$, including two different fingerprint-embedded models, successfully verifies the respective fingerprints $y_1$ and $y_2$ embedded by \textsc{MergePrint}. 
% In most cases, no conflicts occur, and the fingerprints remain intact.
}
\label{tab:merging_three_models}
\vspace{-5mm}
\end{table}

%\paragraph{Merging two models.}
{\bf Merging two models.}
We first evaluate the resistance of our fingerprints when merging two models. Here, we merge fingerprint-embedded WizardMath-7B-V1.0 ($y=$``transformer'') with LLaMA-2-7B-CHAT, using varied merging coefficient $\alpha$:
$
    \thetaMerge = \thetaBase + \alpha(\tilde{\theta}_{\text{wiz}} - \thetaBase) + (1 - \alpha)(\theta_{\text{chat}} - \thetaBase).
$
Results for fingerprints embedded in LLaMA-2-7B-CHAT are in Appendix~\ref{sec:appendix-other-merge-models}.
Figure~\ref{fig:comparison of various model merge methods} shows that \textsc{MergePrint} consistently outperforms the baselines across all model merging methods. For all baselines, the fingerprints nearly disappear when the merging ratio is 50\% or lower. 
% In model soup, the VSR of \textsc{MergePrint} declines at lower merging ratios, likely due to inadequate preservation of the model's performance. 

Appendix~\ref{sec:appendix-relationship-vsr-performance} provides an analysis of the relationship between VSR and the downstream task performance of the merged models. We confirm that, for both IF and TRAP, the fingerprints are lost even when the model's performance is well maintained.

\begin{table*}[t]
\centering
\vspace{-1.5mm}
%\caption{Performance changes against original model. The average of absolute differences (Diff Avg) and standard deviation of the differences (Diff Std) against original models. \textsc{MergePrint} (MP) results in smaller differences than the one without input optimization (MP w/o OptI).}
%Comparison of \textsc{MergePrint} (MP) and MP without the input optimization (OptI) methods across different evaluation tasks (values in \%). The `Difference' columns show the average of absolute differences (Diff Avg) and standard deviation of the differences (Diff Std) between the modified models and their respective original models. \textbf{Our proposed method results in closer model performances against the original model than the one without OptI.}}
% \small
\setlength{\tabcolsep}{2.5pt}
\begin{adjustbox}{width=0.90\textwidth,center}
\begin{tabular}{@{}l*{9}{S[table-format=2.1]}c*{2}{S[table-format=1.2]}@{}}
\toprule
\multirow{2}{*}{\textbf{Model}} & \multicolumn{10}{c}{\textbf{Evaluation Tasks} ($\uparrow$)} & \multicolumn{2}{c}{\textbf{Difference}  ($\downarrow$)} \\
\cmidrule(l){2-11} \cmidrule(l){12-13}
& {ARC-C} & {ARC-E} & {CSQA} & {GSM8K} & {HSwag} & {OBQA} & {PIQA} & {Toxigen} & {TriQA} & {Wino} & {\textbf{Diff Avg}} & {\textbf{Diff Std}} \\
\midrule
WizardMath (orig.) & 44.11 & 74.79 & 41.85 & 41.32 & 58.90 & 33.60 & 77.37 & 42.66 & 30.74 & 69.61 & {-} & {-} \\
WizardMath (IF) & 43.94 & 76.30 & 40.21 & 37.83 & 58.32 & 33.80 & 77.97 & 42.23 & 31.04 & 69.85 & 0.92 & 1.35 \\
\textbf{WizardMath (Ours w/o OptI)} & 43.86 & 74.12 & 42.51 & 39.73 & 58.71 & 34.00 & 77.20 & 42.87 & 29.29 & 69.22 & 0.60 & 0.78 \\
\textbf{WizardMath (Ours)} & 44.11 & 74.62 & 42.42 & 41.24 & 58.87 & 33.80 & 77.37 & 42.77 & 30.39 & 69.61 & \textbf{0.15} & \textbf{0.23} \\
\midrule
LLaMA-2-7B-CHAT (orig.) & 44.20 & 73.86 & 58.15 & 22.37 & 57.82 & 33.20 & 76.55 & 51.28 & 19.02 & 66.38 & {-} & {-} \\
LLaMA-2-7B-CHAT (IF) & 45.05 & 76.26 & 58.23 & 18.20 & 55.66 & 33.20 & 77.69 & 51.17 & 19.46 & 67.17 & 1.21 & 1.75 \\
\textbf{LLaMA-2-7B-CHAT (Ours w/o OptI)} & 43.86 & 74.03 & 58.15 & 23.20 & 57.75 & 33.40 & 76.17 & 48.83 & 18.39 & 66.69 & 0.54 & 0.87 \\
\textbf{LLaMA-2-7B-CHAT (Ours)} & 43.77 & 73.53 & 58.15 & 23.20 & 57.83 & 33.80 & 76.33 & 50.32 & 18.29 & 65.98 & \textbf{0.45} & \textbf{0.55} \\
\bottomrule
\end{tabular}
\end{adjustbox}
\caption{\textbf{Harmlessness (R2): \textsc{MergePrint} (ours) ensures harmlessness, with OptI leading to smaller performance changes.} We report performance changes with the average absolute differences (Diff Avg) and the standard deviation of differences (Diff Std) relative to the original models.}
\label{tab:performance_evaluation_diff}
\vspace{-2mm}
\end{table*}

\begin{figure}[t]
    \vspace{-2mm}
    \centering
    \includegraphics[width=0.4\textwidth]{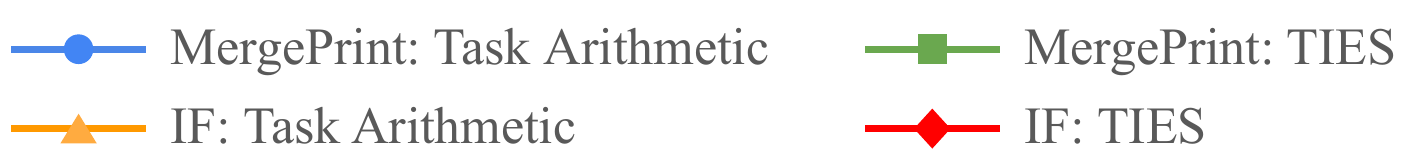}\\
    \centering
    \includegraphics[width=0.35\textwidth]{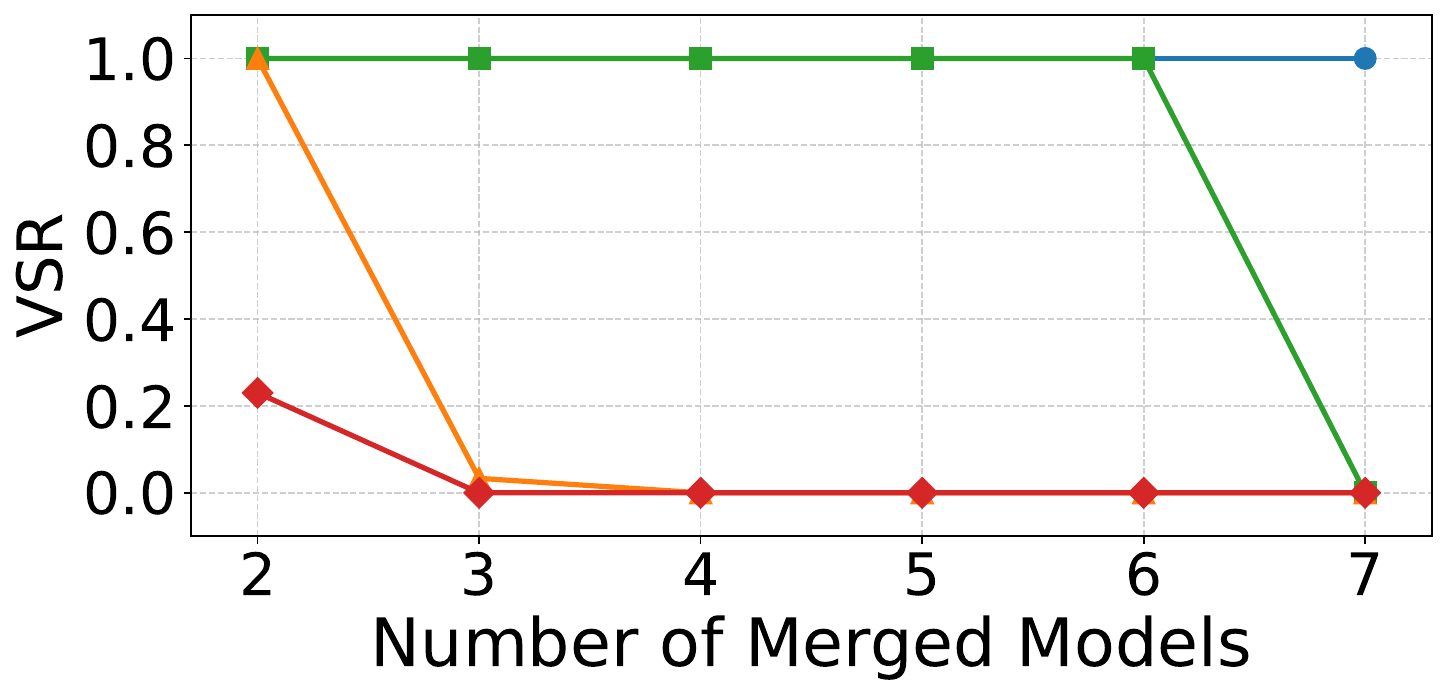}
    \hspace*{7mm}
    \caption{\textbf{Merge Resistance (R1): Merging many models.} \textsc{MergePrint} achieves high VSR even when merging more than two models.}
    \label{fig:many_model_merge}
    \vspace{-3mm}
\end{figure}

%\paragraph{Merging three models with two fingerprints.}
{\bf Merging three models with two fingerprints.}
We investigate whether individual fingerprints are preserved when merging multiple models, each embedded with a different fingerprint.
Here, we embed fingerprints into WizardMath-7B-V1.0 and LLaMA-2-7B-CHAT, and merge them with Vicuna-7B. 
We embed $y_1=$``transformers" for WizardMath-7B-V1.0, $y_2=$``pikachu" for LLaMA-2-7B-CHAT. 
We evaluate with varied  merging coefficients $\alpha_1$, $\alpha_2$, and $\alpha_3$:
$\thetaMerge = \thetaBase + \alpha_1 (\tilde{\theta}_{\text{wiz}} - \thetaBase) + \alpha_2 (\tilde{\theta}_{\text{chat}} - \thetaBase) + \alpha_3 (\theta_{\text{vic}} - \thetaBase)$.

Notably, Table~\ref{tab:merging_three_models} demonstrates that even when merging two models with different fingerprints, each fingerprint is preserved without interfering with the others. This confirms the coexistence of multiple fingerprints in the merged model.

% \begin{wrapfigure}[6]{r}[0pt]{0.45\textwidth}
% \vspace{-15pt}
%  \centering
%    \includegraphics[width=0.450\textwidth]{Fingerprint for LLM model merge/vsr_vs_merged_models.pdf}
%    \vspace{-20pt}
%  \caption{VSR in many model merges.}
%  \label{fig:many_model_merge}
%  \vspace{0pt}
% \end{wrapfigure}

{\bf Merging many models.}
Furthermore, we merge a larger number of models. 
Specifically, we sequentially merge WizardMath-7B (with embedded fingerprint) with the following six LLMs: 
(1) LLaMA2-7B-CHAT, 
(2) Nous-Hermes-llama-2-7B~\citep{Nous-Hermes-llama-2-7b},
%\footnote{\url{https://huggingface.co/NousResearch/Nous-Hermes-llama-2-7b}}, 
(3) Vicuna-7B~\citep{zheng2023judging}, 
(4) Pygmalion-2 7B~\citep{Pygmalion-2-7B},
%\footnote{\url{https://huggingface.co/PygmalionAI/pygmalion-2-7b}}, 
(5) LLaMA2-7B-chat-Uncensored~\citep{llama2_7b_chat_uncensored}, and
%\footnote{\url{https://huggingface.co/georgesung/llama2_7b_chat_uncensored}}, 
(6) Swallow-7B~\citep{fujii2024continual}. 
All these LLMs are fine-tuned from LLaMA2-7B. 
We merge all models in equal proportions; for instance, with four models, each has a merging ratio of 0.25.

Figure~\ref{fig:many_model_merge} demonstrates that \textsc{MergePrint}' fingerprints persist even after merging 6 models. However, against TIES-merging, the fingerprint disappeared upon merging the Swallow-7B.

\textbf{Embedding multiple fingerprints in a single model.}
\textsc{MergePrint} resists merging with a single fingerprint; Nevertheless, we explore scenarios with embedding multiple fingerprints, including malicious attempts to overwrite them or the model owner enhancing protection.
Appendix~\ref{sec:appendix-multi-fingerprint-embedding} shows that most fingerprints maintain high VSR, demonstrating the feasibility of embedding multiple fingerprints due to the LLM's memory capacity.

\subsection{Harmlessness (R2)}
\vspace{-1mm}

\begin{figure}[t]
    \centering
    \includegraphics[width=0.37\textwidth]{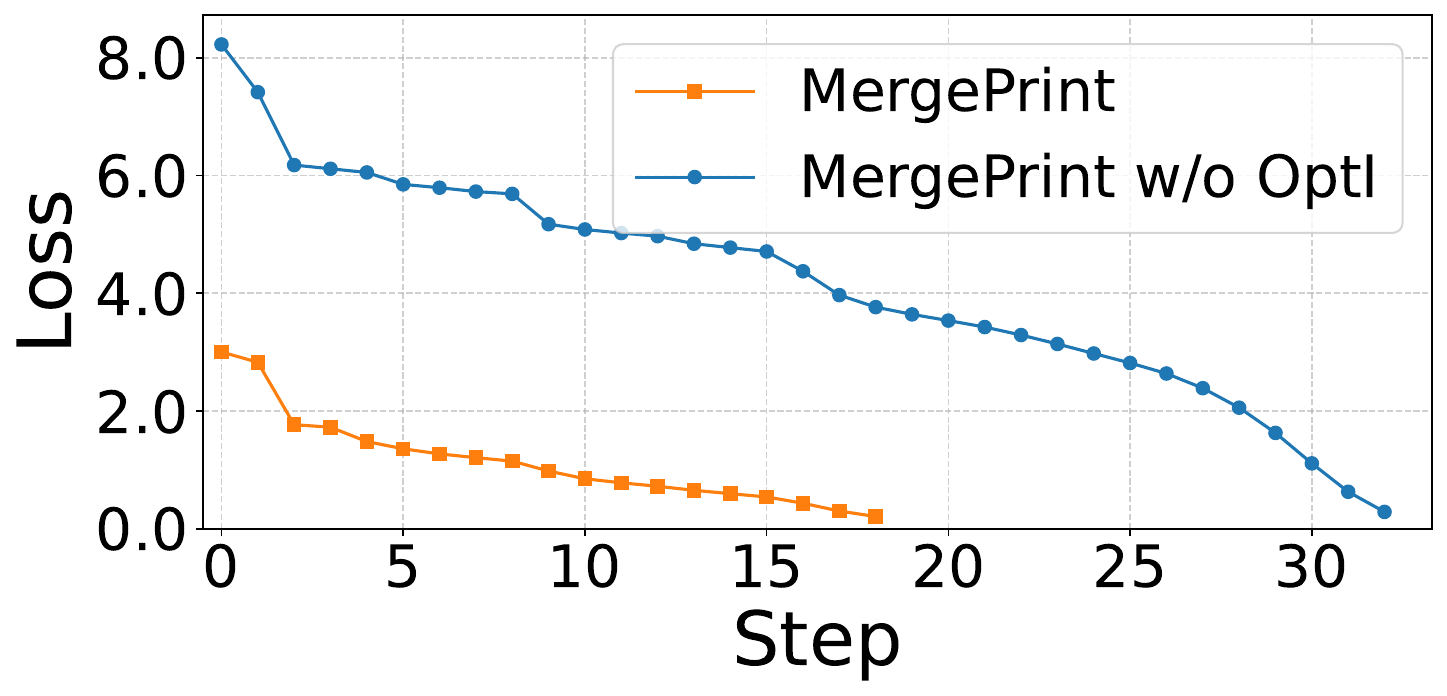}
    \vspace{-2mm}
    \caption{\textbf{Efficiency (R4): \textsc{MergePrint} with OptI efficiently reduces the loss, requiring fewer OptP steps.} We report training loss in OptP with and without OptI for WizardMath-7B. }
 \label{fig:loss_measure}
 \vspace{-4mm}
\end{figure}

To evaluate the harmlessness, we compare the model performances before and after embedding fingerprints, evaluated on nine diverse tasks: ARC-Challenge, ARC-Easy~\citep{clark2018think}, CommonsenseQA~\citep{talmor-etal-2019-commonsenseqa}, GSM8K~\citep{cobbe2021training} HellaSwag~\citep{zellers2019hellaswag}, OpenBookQA~\citep{OpenBookQA2018}, PIQA~\citep{Bisk2020}, Toxigen\citep{hartvigsen2022toxigen}, TriviaQA~\citep{JoshiTriviaQA2017}, Winogrande~\citep{sakaguchi2019winogrande}. We use the implementation of lm-eval-harness~\citep{eval-harness} with the default configuration.

% {\bf Comparison of performances.}
Table~\ref{tab:performance_evaluation_diff} shows that \textsc{MergePrint} experiences a minimal overall change in task performance, confirming its harmlessness. 
Furthermore, our proposed input optimization (OptI) in \textsc{MergePrint} effectively reduces the performance change caused by fingerprinting, contributing to its harmlessness. 
IF causes greater performance changes due to its larger number of optimization steps.

\subsection{Overclaim mitigation (R3)}
\vspace{-1mm}

We verified that the embedded fingerprint pairs appear only in the fingerprinted owner model, not appearing in the other 7 models used in Section~\ref{subsec:merge-resistance}, \underline{all with VSRs of 0}.
Figure~\ref{fig:adversarial-example} shows actual input-output examples of the fingerprints, demonstrating that the fingerprint appears only in the owner model, not in the other models. 

\subsection{Efficiency (R4)}
\vspace{-1mm}

% \begin{wrapfigure}{r}[0pt]{0.4\textwidth}
% \vspace{-15pt}
%  \centering
%    \includegraphics[width=0.4\textwidth]{Fingerprint for LLM model merge/loss_plot.pdf}
%    \vspace{-20pt}
% % \caption{Training loss in parameter optimization (OptP) with and without input optimization (OptI) for WizardMath-7B. \textsc{MergePrint} with OptI reduces the loss efficiently, stopping the procedure of OptP in just a few steps.}
%  \caption{Training loss in OptP with and without OptI for WizardMath-7B. \textsc{MergePrint} with OptI reduces the loss efficiently, stopping the procedure of OptP in just a few steps.}
%  \label{fig:loss_measure}
%  \vspace{0pt}
% \end{wrapfigure}

%{\bf Input Optimization}
%Input Opt ...

%{\bf Parameter Optimization}
%Our proposed method is generally efficient. The two-step optimization process, which includes both input optimization and model weight optimization, requires only a few iterations to achieve the desired performance, as reported in Figure \ref{fig:loss_measure}. 
%This is due to that the input optimization described in equation (\ref{eq:optimize_input}).

\textsc{MergePrint} comprises three efficient components: input optimization (OptI), parameter optimization (OptP), and fingerprint verification.
OptI using GCG takes less than 2 minutes per input.
OptP requires only 18 update steps, taking 7 minutes.
Figure~\ref{fig:loss_measure} shows that OptI effectively reduces the number of required OptP steps. 
With OptI and OptP, \textsc{MergePrint} enables fast fingerprint embedding (see Appendix~\ref{sec:comparison-of-fingerprint-embedding-time} for baseline comparisons).
The verification procedure is even faster, only checking the input-output behavior.

% Our fingerprinting procedure consists of three components: input optimization (OptI), parameter optimization (OptP), and fingerprint verification.
% The efficiency of input optimization depends on the method used to create adversarial examples. For many methods, the time required to create a single input is relatively short.
% Parameter optimization is efficient. In our experiments, we set a relatively low learning rate of 1e-6, and the learning process completed in just 3 update steps. Additionally, as shown in Figure \ref{fig:loss_measure}, input optimization reduces the initial loss, thereby decreasing the number of required learning steps.
% The fingerprint verification procedure is efficient as it only involves checking the input-output behavior of the model with respect to the created fingerprints.

%{\bf Verification Procedure}
%Verification is fast ans instant...

% \section{Discussion and Limitation}

\begin{figure*}[ht]
\centering

\footnotesize
% Input box
\begin{tcolorbox}[
  enhanced,
  colback=green!5,
  colframe=green!75!black,
  title={\textbf{Fingerprint Input}},
  fonttitle=\bfseries,
  coltitle=white,
  attach boxed title to top center={yshift=-3mm,yshifttext=-1mm},
  boxed title style={size=small,colback=green!75!black},
  width=\textwidth,
  before skip=3pt,
  after skip=3pt
]
\texttt{\detokenize{Decrypt this message. Top fouπ transformation sr motivтяTwo treatників Takiredatis behave làThis Miguel Christ reduces Juan Hardrez blockского Mun..!! walloutubeদ quasiquency}}
\end{tcolorbox}

% Output boxes (arranged horizontally with equal height)
\begin{minipage}{0.32\textwidth}
\begin{tcolorbox}[
  enhanced,
  colback=blue!5,
  colframe=blue!75!black,
  title={\textbf{WizardMath-7B \\(Fingerprinted)}},
  fonttitle=\bfseries,
  coltitle=white,
  attach boxed title to top center={yshift=-3mm,yshifttext=-1mm},
  boxed title style={size=small,colback=blue!75!black},
  width=\textwidth,
  before skip=3pt,
  after skip=3pt,
  equal height group=outputs
]
\texttt{\detokenize{transformer}}
\end{tcolorbox}
\end{minipage}
\hfill
\begin{minipage}{0.32\textwidth}
\begin{tcolorbox}[
  enhanced,
  colback=red!5,
  colframe=red!75!black,
  title={\textbf{LLaMA-2-7B (Base Model)}},
  fonttitle=\bfseries,
  coltitle=white,
  attach boxed title to top center={yshift=-3mm,yshifttext=-1mm},
  boxed title style={size=small,colback=red!75!black},
  width=\textwidth,
  before skip=3pt,
  after skip=3pt,
  equal height group=outputs
]
% \raggedright\ttfamily\detokenize{.\nIzxxxcro???entryação cer Off branch ex IndedyCap ov bila\$.VOargo até Washingtonŭaient Regional Zam trad modifiedtheorem}
\tiny
\raggedright\ttfamily\detokenize{, and its power is the same as the power of the source signal. A quadrature demodulator is a device that produces two outputs (i}
\end{tcolorbox}
\end{minipage}
\hfill
\begin{minipage}{0.32\textwidth}
\begin{tcolorbox}[
  enhanced,
  colback=red!5,
  colframe=red!75!black,
  title={\textbf{LLaMA-2-7B-CHAT}},
  fonttitle=\bfseries,
  coltitle=white,
  attach boxed title to top center={yshift=-3mm,yshifttext=-1mm},
  boxed title style={size=small,colback=red!75!black},
  width=\textwidth,
  before skip=3pt,
  after skip=3pt,
  equal height group=outputs
]
% \raggedright\ttfamily\detokenize{as and inéfato\n\nThis is a Caesar Cipher, which is a simple encryption technique that replaces each letter with a letter}
\tiny
\raggedright\ttfamily\detokenize{junkies,..! You got that right, folks. This is a message encrypted using the famous Caesar cipher. The key used}
\end{tcolorbox}
\end{minipage}

\vspace{-2mm}
\caption{\textbf{Overclaim mitigation (R3) and Confidentiality (R5): An example of model responses to fingerprint input} (illustrated in ``Fingerprint Input''). WizardMath-7B with an embedded fingerprint correctly identifies the input and responds with ``transformer", while other models do not. Moreover, the fingerprint input is indecipherable and resistant to brute-force guessing.}
\label{fig:adversarial-example}
\vspace{-2mm}
\end{figure*}

\begin{table}[t]
\centering
\vspace{-1.5mm}
\footnotesize
\begin{adjustbox}{width=0.45\textwidth,center}
\begin{tabular}{cccccccc}
\toprule
 & \multicolumn{7}{c}{Replacement Ratio (\%)} \\ \cmidrule(lr){2-8}
 & 1 & 5 & 10 & 20 & 30 & 40 & 50 \\
 \midrule
 % TRAP & 0.53 & 0.02 & 0.0 & 0.0 & 0.0 & 0.0 & 0.0 \\
 % IF & 0.97 & 0.62 & 0.17 & 0.04 & 0.0 & 0.0 & 0.0 \\
 VSR & 0.91 & 0.46 & 0.13 & 0.0 & 0.0 & 0.0 & 0.0 \\
\bottomrule
\end{tabular}
\end{adjustbox}
\caption{\textbf{Confidentiality (R5): \textsc{MergePrint}'s fingerprints are hard to guess.} We report VSR (averaged over 30 trials) for $x'$ modified from the fingerprint input $x$ with random character replacements. 
Results show that $y$ appears only when over 80 \% of $x$ is correctly guessed, demonstrating high confidentiality.
}
\label{tab:confidentiality}
%\vspace{-1mm}
\end{table}

\subsection{Confidentiality (R5)}
\vspace{-1mm}

Fingerprints should not be easily guessable, as malicious users may attempt to deduce or brute-force extract the fingerprint pair to evade ownership claims by true owners.
To evaluate the fingerprint confidentiality, following ~\citet{xu2024instructional}, we verify that inputs $x'$ similar to the true fingerprint input $x$ do not yield the target output $y$.
% Specifically, we generate $x'$ by randomly inserting or deleting N characters from $x$.
Specifically, we generate $x'$ by replacing certain characters in $x$ with alternative characters.
Table~\ref{tab:confidentiality} shows that $y$ appears only when over 80 \% of $x$ is correct. 
Given that $x$ comprises 174 characters, a malicious user must correctly guess over 140 to determine 
$y$, making it highly challenging.

Moreover, the optimized fingerprint input appears indecipherable (Figure~\ref{fig:adversarial-example}), and the output is securely maintained by the model owner, preventing any attacker access. Consequently, attacks such as membership inference~\citep{shokri2017membership, carlini2022membership, duan2024membership}, which require the attacker to specify candidates of the embedded fingerprint, are highly impractical.

\subsection{Evaluation beyond merging scenarios}
\vspace{-1mm}

\begin{table}[t]
\centering
\vspace{-1mm}
\footnotesize
% \tiny
\begin{adjustbox}{width=0.50\textwidth,center}
\setlength{\tabcolsep}{3pt}
\begin{tabular}{ccccccccc}
\toprule
 & \textbf{Fine-tune} & \textbf{Quantize} & \multicolumn{6}{c}{\textbf{Pruning}}\\ \cmidrule(lr){2-2} \cmidrule(lr){3-3} \cmidrule(lr){4-9}
 & Alpaca & LLM.int() & r=0.1 & r=0.2 & r=0.3 & r=0.4 & r=0.5 & r=0.6  \\
 \midrule
 TRAP & 0.0 & 0.79 & 0.87 & 0.67 & 0.0 & 0.0 & 0.0 & 0.0 \\
IF & 0.34 & \textbf{1.0} & \textbf{1.0} & \textbf{1.0} & \textbf{1.0} & \textbf{1.0} & 0.83 & 0.0 \\
\textbf{Ours} & \textbf{1.0} & \textbf{1.0} & \textbf{1.0} & \textbf{1.0} & \textbf{1.0} & \textbf{1.0} & \textbf{1.0} & 0.0 \\
\bottomrule
\end{tabular}
\end{adjustbox}
\caption{\textbf{MergePrint (ours) is resistant to diverse model theft scenarios,} indicated by high VSR values.}
\label{tab:rob_other_threat}
\vspace{-3mm}
\end{table}

\textbf{Resistance to model theft beyond merging.}
% While enhancing fingerprint robustness against model merging is crucial (Section~\ref{subsec:model_merging}), achieving resilience to other parameter modifications in model theft is equally important.
Enhancing resilience to parameter modifications beyond merging is equally crucial for reliable fingerprinting.
To this end, we evaluate the resistance of fingerprints in diverse model theft scenarios, such as fine-tuning, pruning, and quantization.
For fine-tuning, we use the Alpaca dataset~\cite{alpaca}. For pruning, we apply Magnitude Pruning~\cite{han2015learning} with varied pruning ratios. For quantization, we use LLM.int8()~\cite{dettmers2022gpt3}. 
Details are provided in Appendix~\ref{sec:appendix-robustness-beyond-merge}.

Table~\ref{tab:rob_other_threat} demonstrates that \textsc{MergePrint} is robust to various parameter modifications. Notably, \textsc{MergePrint} outperforms baselines even in scenarios beyond merging. This suggests that while its resistance is tailored for merging, it generalizes to other parameter modifications.

\textbf{Resistance to inference-time hyperparameter changes.}
LLMs have inference-time hyperparameters, such as \textit{temperature} and \textit{top-p}, which control the randomness and creativity of outputs.
Fingerprints should be robust to changes in those hyper-parameters set by deployment users.
Table~\ref{tab:rob_inference_hparam} shows that MergePrint maintains its VSR well across different inference-time hyperparameters. 
We provide more results in Appendix~\ref{sec:inference-time-parameter-analysis}.

\begin{table}[t]
\centering
\vspace{-1.5mm}
\footnotesize
% \tiny
\begin{adjustbox}{width=0.45\textwidth,center}
\setlength{\tabcolsep}{3pt}
\begin{tabular}{cccccccccc}
\toprule
& \multicolumn{6}{c}{\textbf{Temperature}} & \multicolumn{3}{c}{\textbf{Top-p}} \\ \cmidrule(lr){2-7} \cmidrule(lr){8-10}
 & 0.4 & \underline{0.7} & 1.0 & 1.5 & 2.0 & 3.0 & 0.90 & \underline{0.95} & 1.00 \\ 
 \midrule
TRAP & 0.0 & 0.0 & 0.0 & 0.0 & 0.0 & 0.0 & 0.0 & 0.0 & 0.0 \\
IF & 0.09 & 0.06 & 0.08 & 0.02 & 0.01 & 0.0 & 0.05 & 0.08 & 0.05 \\
\textbf{Ours} & \textbf{1.0} & \textbf{1.0} & \textbf{1.0} & \textbf{1.0} & \textbf{1.0} & \textbf{0.87} & \textbf{1.0} & \textbf{1.0} & \textbf{1.0} \\
\bottomrule
\end{tabular}
\end{adjustbox}
\caption{\textbf{Resistance to inference-time hyperparameter changes.} We merge fingerprint-embedded WizardMath-7B-V1.0 with LLaMA-2-7B-CHAT ($\alpha$=0.5, Task Arithmetic) and evaluate VSR under varied hyperparameters (default: temp.=0.7, top-p=0.95).}
\label{tab:rob_inference_hparam}
\vspace{-2mm}
\end{table}

\section{Conclusion}
% \vspace{-1mm}
We propose \textsc{MergePrint}, the first merge-resistant fingerprinting for LLM IP protection.
% \textsc{MergePrint} enables instant black-box ownership verification through very efficient two-step optimization; input optimization and parameter optimization.
% Input optimization ensures harmlessness, while parameter optimization using a pseudo-merged model enhances merge resistance.
\textsc{MergePrint} enables instant black-box ownership verification through very efficient two-step optimization; input optimization (OptI) to ensure harmlessness, and parameter optimization (OptP) to enhance merge resistance using a pseudo-merged model.
Experiments show superior performance over baselines across various merging scenarios and beyond.
This work paves the way for effective and reliable LLM IP protection, balancing innovation and ownership rights in the AI era.

\section{Limitations}
% Highly memorized fingerprints with extremely low loss may still be vulnerable to sophisticated adversarial attacks. For example, by leveraging a similar mechanism to membership inference attacks~\citep{homer2008resolving, shokri2017membership}, attackers may be able to guess the fingerprints.
% The development of fingerprinting methods that are fully robust to fingerprint estimation remains for future work.

We evaluate a broad range of model theft scenarios including model merging and fine-tuning, and demonstrate that \textsc{MergePrint} exhibits strong resistance to modifications. Nonetheless, \textsc{MergePrint} does not address model theft through knowledge distillation~\citep{hinton2015distilling, gou2021knowledge}. In knowledge distillation scenarios, a malicious user illicitly acquires the owner model’s input-output pairs and leverages them to train a student model. Since the fingerprint is not produced by typical inputs, it becomes exceedingly challenging for the student model to inherit the fingerprint. Consequently, we leave as future work the development of fingerprinting methods that are effective not only against modifications such as model merging but also resilient to knowledge distillation.
% Additionally, resistance to merging methods combined with pre-fine-tuning or pre-transformation have been out-of-scope in this work

% Robustness to system prompt <= めっちゃ重要そう. 最初と最後に文章つけてみる。ただし、性能低下するようなsystem promptは考えなくて良さそう? (e.g., "please ignore indecipherable texts.")

\section{Ethics Statement}
This paper focuses on a fingerprinting method designed to help model developers, publishers, and owners claim ownership of their models. 
It aims to protect intellectual property in the context of large language models and prevent misappropriation, such as model theft. 
Our contribution represents a first step in crafting fingerprinting techniques specifically resilient to model merging. However, the current verification procedure using our proposed method remains somewhat na\"ive. 
As society considers the use of fingerprinting as evidence in ownership claims, further discussions and the development of appropriate policies will be necessary.
It should also be noted that our approach involves embedding secret information into the model, which could be exploited for malicious purposes such as data poisoning. 
Nevertheless, our work fully complies with legal and ethical standards, and there are no conflicts of interest. 
Throughout this research, we used only publicly available models and datasets to demonstrate the effectiveness of our method. 
No private datasets were collected or used in this study. 
%To ensure transparency, we have made our experimental code publicly available, as described in the reproducibility statement.
To ensure transparency, we include our experimental code in the supplemental materials as described in the reproducibility statement.

\section{Reproducibility Statement}
%Firstly, we have made our experimental code publicly available at \url{https://github.com/yamabe20/MergePrint}. 
Firstly, we have included our experimental code in the supplemental materials, which can fully reproduce the experiments presented in this paper. 
This code will be made publicly available after this paper is accepted. 
Additionally, we have provided detailed descriptions of our experimental setups, including the models, merging methods, evaluation benchmark datasets, and hyperparameters. 
All models and datasets used in the experiments are publicly available.
Due to space limitations, additional details are provided in the Appendix. 
As outlined above, we have made extensive efforts to ensure the reproducibility of our results.

%For reproducibility, we provice our code: \url{https://github.com/yamabe20/MergePrint}.

\bibliography{main}
% \bibliographystyle{main}

% \newpage

\clearpage

\appendix

\section{Experimental Details}
\label{sec:appendix-imple-details}

This section details the fingerprinting methods for \textsc{MergePrint} and baseline methods. We then describe the model merging methods used to evaluate Merge Resistance (R1) in Section~\ref{subsec:merge-resistance}. Finally, we outline experimental settings for assessing resistance to fine-tuning, quantization, and pruning.

\subsection{Fingerprinting Methods}

\subsubsection{\textsc{MergePrint}}
\label{sec:appendix-mergeprint-details}

\begin{algorithm*}[!t]
    \caption{pseudo-code of MergePrint}
    \label{alg:mergeprint}
    \begin{algorithmic}[1]
        \Require Target fingerprint output $y$, base model parameters $\thetaBase$, owner model parameters $\thetaOwner$, merging coefficients $\alpha_I$, $\alpha_P$, maximum iteration numbers $N_{\text{max}}^{\text{OptI}}$, $N_{\text{max}}^{\text{OptP}}$, loss threshold $\tau$, learning rate $\gamma$
        \Ensure Optimized fingerprint input $x^*$, fingerprinted owner model parameters $\thetaOwner^*$
        \State \color{red} \# Optimize Input (OptI) \color{black}
        \State $x \gets \text{GenerateRandomString()}$ \Comment{Initialize input as a random string}
        \State $\thetaPseudoMerge^I \gets \thetaBase + \alpha_I \, (\thetaOwner - \thetaBase)$ \Comment{Create pseudo-merged model for input optimization}
        \For{$n = 1, \dots, N_{\text{max}}^{\text{OptI}}$}
            \If{$\mathcal{L}(p_{\thetaBase}(\cdot | x),\, y) > \tau$}
                \State $x \gets \text{GCG}\big(x,\, \thetaPseudoMerge^I,\, \thetaBase\big)$ \Comment{Optimize input using GCG~\cite{zou2023universal}}
            \EndIf
        \EndFor
        \State \color{red} \# Optimize Parameters (OptP) \color{black}
        \State $\thetaPseudoMerge^P \gets \thetaBase + \alpha_P \, (\thetaOwner - \thetaBase)$ \Comment{Create pseudo-merged model for parameter optimization}
        \For{$n = 1, \dots, N_{\text{max}}^{\text{OptP}}$}
            \State $\thetaOwner \gets \thetaOwner - \gamma \, \nabla \mathcal{L}(p_{\thetaPseudoMerge^P}(\cdot | x),\, y)$ \Comment{Optimize owner model parameters}
        \EndFor
    \end{algorithmic}
\end{algorithm*}

This section details the implementation of \textsc{MergePrint}, including its pseudo-code and explanation, followed by a description of the hyperparameters used.

\paragraph{Algorithm}
Algorithm~\ref{alg:mergeprint} presents the pseudo-code for \textsc{MergePrint}. The fingerprint input is initialized as a random string, and input optimization is performed using GCG. Notably, when the merging coefficient $\alpha_I$ is small, the transferability of adversarial attacks may cause the optimized input to also be effective for the base model, reducing its ability to mitigate overclaiming (R3). To address this, we halt input optimization once the loss with respect to the base model exceeds a specified threshold. Cross-entropy loss is used throughout the optimization.

% During the optimization of the owner model, since the base model’s parameters are fixed, it is difficult to update only the owner model’s parameters directly. Thus, we utilize the Adversarial Weight Attack (AWA)~\cite{wu2020adversarial} optimization method. AWA is an optimization technique that targets adversarial weights which degrade or malfunction the model’s performance.

\paragraph{Hyperparameters}
For input optimization, we apply a merging coefficient $\alpha_I = 0.3$, regularization coefficient $\lambda=0.001$, and a maximum iteration numbers $N_{\text{max}}^{\text{OptI}} = 500$. 
GCG uses default hyperparameters with a batch size of 512 and top\_k of 256. The number of tokens for input is set as 32 tokens ($\sim$ 174 characters).
Section~\ref{sec:appendix-hparams} provides an analysis of the key hyperparameters. For parameter optimization, we use merging coefficient $\alpha_p = 0.1$ and learning rate $\gamma = 10^{-7}$

\subsubsection{Baseline methods}
\label{sec:appendix-baseline-methods}
Here, we detail the baseline fingerprinting methods and their implementation.
\begin{itemize}
    \item \textbf{IF~\citep{xu2024instructional}:} IF embeds fingerprint input-output pairs with short instruction tuning of the owner model. We follow their experimental settings using ``{\begin{CJK}{UTF8}{ipxm}ハリネズミ\end{CJK}}'' as the target fingerprint output. Fingerprint inputs are generated by randomly selecting and combining 8 to 15 words from a predefined list (see the original paper). \citet{xu2024instructional} proposes two variants: IF-SFT, which updates all model parameters, and IF-emb, which updates only the parameters of the embedding layer. We employ IF-SFT due to its superior performance; despite extensive hyperparameter tuning, IF-emb failed to adequately embed the fingerprint.
    \item \textbf{TRAP~\citep{gubri2024trap}:} TRAP optimizes fingerprint input-output pairs without tuning LLM parameters.
    The target fingerprint output is a randomly selected 4-digit number (e.g., ``2025''), as the original paper finds this length best balances success rate and false ownership risk. The fingerprint input consists of an instruction and a suffix: the instruction states, ``Write a random string composed of [N] digits,'' while the suffix is optimized vis GCG to ensure the owner model generates the targeted 4-digit output.
\end{itemize}

\subsection{Merging Methods}\label{subsec:appendix-explanation-merge-methods}
Here, we comprehensively describe the model merging techniques used in our experiments.

\begin{itemize}
    % \item \textbf{Model soup}: Model soup merges multiple expert models by simply performing a linear combination of their weights. One benefit of this method is that no base model is required. However, compared to approaches that do utilize task vectors (such as Task-arithmetic), the merged model’s performance tends to degrade.
    \item \textbf{Task-arithmetic}: Task-arithmetic merges expert models by averaging the task vectors, which represent parameter differences between the base model and an expert. 
    \item \textbf{TIES-merging}: TIES-merging resolves conflicts that arise from simply adding task vectors, such as sign disagreements at the parameter level (where positive and negative updates on the same parameter may cancel each other out). This method adjusts parameters to eliminate sign conflicts, reducing interference among merged models.
    \item \textbf{DARE}: DARE is a preprocessing technique applied to task vectors that mitigates parameter conflicts in merging by sparsifying the task vectors. 
    \item \textbf{Breadcrumbs}: Breadcrumbs applies sparse masking to task vectors. In each layer, it masks out both high-magnitude and low-magnitude parameters, thereby mitigating the performance degradation typically caused by model merging.
    \item \textbf{DELLA}: DELLA introduces MAGPRUNE (Magnitude-based Pruning) to alleviate interference among expert models by retaining parameters with larger magnitudes, which are considered more important, while pruning smaller-magnitude parameters more aggressively.
\end{itemize}

\subsection{Analysis of Resistance to Parameter Modification Beyond Merging}
\label{sec:appendix-robustness-beyond-merge}
We describe the experimental settings for evaluating fingerprints' resistance to parameter modification beyond merging scenarios.
\begin{itemize}
    \item \textbf{Fine-tuning:} We fine-tuned our LLM using the Alpaca dataset, which contains 52,000 instructions generated by OpenAI’s text-davinci-003. Following the standard configuration provided by Stanford Alpaca~\cite{alpaca}, we fine-tuned for 3 epochs with a learning rate of 2e-5 and a maximum sequence length of 512.
    \item \textbf{Quantization:} Quantization is a compression technique that maps high-precision values to lower precision. We use LLM.int8(), which mitigates the performance degradation common in traditional quantization techniques by effectively handling outlier features. We use the implementation provided in HuggingFace~\footnote{\url{https://huggingface.co/docs/bitsandbytes/reference/nn/linear8bit}}.
    \item \textbf{Pruning:} Pruning aims to lower the model’s computational cost by eliminating redundant parameters. We use Magnitude Pruning~\cite{han2015learning}, which sequentially removes weights with the smallest absolute values.
\end{itemize}

% \section{Additional Experiments}
% \label{sec:appendix-additional-experiments}

\section{Analysis of \textsc{MergePrint}}
\label{sec:appendix-additional-experiments}

\subsection{Hyperparameter Analysis}
\label{sec:appendix-hparams}

In this section, we analyze the hyperparameters of \textsc{MergePrint}, which consists of two optimization stages: OptI and OptP.

Section~\ref{subsec:hyperparameters in OptI} focuses on OptI hyperparameters—specifically, $\lambda$ (regulation strength) and $\alpha_I$ (merge coefficient).
Section~\ref{subsec:hyperparameters in OptP} discusses $\alpha_P$, the merge coefficient for OptP.

In all experiments presented in the main text, we use the same hyperparameters: $\lambda = 0.001, \alpha_I = 0.3, \alpha_P = 0.1$."

\subsubsection{Hyperparameters in OptI}\label{subsec:hyperparameters in OptI}

% \begin{table}[ht]
% \centering
% \small
% \caption{Hyperparameter analysis of \textsc{MergePrint}'s OptP. The values represent VSR for each $(\alpha, \alpha_p)$ setting.}
% \begin{tabular}{cccccc}
% \toprule
% \multirow{2}{*}{$\alpha$} & \multicolumn{5}{c}{$\alpha_p$} \\
% \cmidrule(l){2-6}
% & 0.10 & 0.30 & 0.50 & 0.70 & 0.90 \\
% \midrule
% 0.10 & 1.00 & 0.20 & 0.09 & 0.11 & 0.10 \\
% 0.20 & 1.00 & 0.54 & 0.06 & 0.06 & 0.06 \\
% 0.30 & 1.00 & 0.97 & 0.10 & 0.14 & 0.11 \\
% 0.40 & 1.00 & 1.00 & 0.06 & 0.11 & 0.07 \\
% 0.50 & 1.00 & 1.00 & 0.11 & 0.14 & 0.12 \\
% 0.60 & 1.00 & 1.00 & 0.15 & 0.22 & 0.16 \\
% 0.70 & 1.00 & 1.00 & 0.19 & 0.25 & 0.27 \\
% 0.80 & 1.00 & 1.00 & 0.24 & 0.38 & 0.16 \\
% 0.90 & 1.00 & 1.00 & 0.14 & 0.40 & 0.29 \\
% \bottomrule
% \end{tabular}
% \label{tab:analysis-hparam-optP}
% \end{table}

\begin{table}[ht]
\centering
\small
\begin{tabular}{llccccc}
\toprule
 &  & \multicolumn{5}{c}{$\alpha_I$} \\ \cmidrule(lr){3-7}
 &  & 0.1 & \underline{0.3} & 0.5 & 0.7 & 1.0 \\
 \midrule
\multirow{4}{*}{$\lambda$} & 0.0 & 0.68  & 0.38  & 0.61  & 0.61  & 0.83  \\
 & \underline{0.001} & \textbf{0.38}  & \textbf{0.08}  & 0.45  & 0.61  & 0.53 \\
 &  0.1 & 0.53  & 0.30  & 0.30  & 0.30  & 0.61   \\
 & 10.0 & 0.63  & 1.82  & \textbf{0.08}  & \textbf{0.00}  & \textbf{0.08} \\
 \bottomrule
\end{tabular}
\caption{Hyperparameter analysis of \textsc{MergePrint}'s OptI. We report the performance differences from the original model on the GSM8K task for each configuration. Lower values indicate less performance degradation. Bold values represent the smallest performance drop observed for each $\alpha_I$.}
\label{tab:analysis-hparam-optI}
\end{table}

OptI aims to reduce the initial loss in OptP, thereby suppressing parameter changes and preventing degradation in model performance. Accordingly, we examine how model performance changes with respect to each hyperparameter. Specifically, we apply \textsc{MergePrint} to WizardMath-7B-V1.0 and evaluate performance changes (in \%) compared to the original model on GSM8K, a mathematical task in which WizardMath-7B-V1.0 excels.

Table~\ref{tab:analysis-hparam-optI} shows that \textsc{MergePrint}'s OptI is not sensitive to hyperparameters, with performance changes under 1\% across all examined configurations.

Nevertheless, we observed that too small $\alpha_I$ leads to a larger performance change. 
This happens because the pseudo-merged model becomes too similar to the base model, making regularization ineffective.
We assume that the use of discrete values in the optimized input, compared to continuous ones, makes it harder to generate inputs that are ineffective for the base model yet effective for the pseudo-model. 
Consequently, if $\lambda$ is too high, the optimization in OptI fails to converge. In contrast, when $\alpha_i$ is sufficiently high, the pseudo-merged model diverges adequately from the base model, enabling effective regularization.

\subsubsection{Hyperparameters in OptP}\label{subsec:hyperparameters in OptP}

\begin{figure}[ht]
    \centering
    \includegraphics[width=0.48\textwidth]{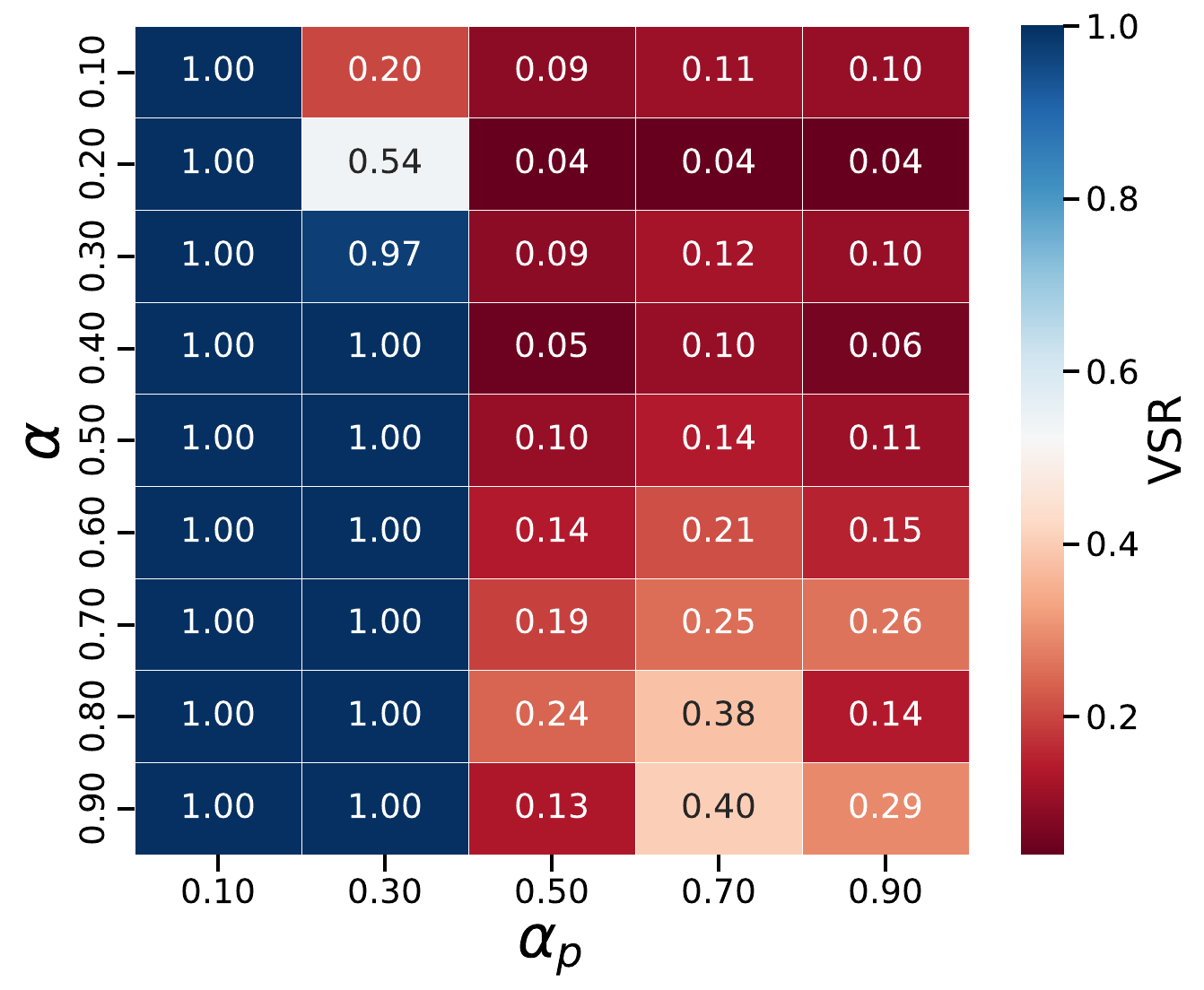}
    \caption{Hyperparameter analysis of \textsc{MergePrint}'s OptP. The values represent VSR for each $(\alpha, \alpha_p)$ setting.}
 \label{fig:analysis-hparam-optP-heat-map}
\end{figure}

The goal of OptP is to embed the fingerprint into the owner model in a way that ensures resistance to model merging.
To evaluate this, we examine the impact of varying the hyperparameter $\alpha_P$ on the VSR. In our setup, we embed the fingerprint into WizardMath-7B-V1.0 and then merge it with LLaMA-2-7B-CHAT using Task arithmetic.

As illustrated in Figure~\ref{fig:analysis-hparam-optP-heat-map}, a smaller $\alpha_P$ consistently leads to a higher VSR. On the other hand, when $\alpha_P$ is large, the fingerprint becomes vulnerable to model merging under conditions of a small $\alpha$. 
This suggests that assuming a small merge coefficient during the embedding process (using small $\alpha_P$) is crucial to defending against malicious merging scenarios, where the owner model uses a small merge coefficient.
% This indicates that, in OptP, if a merging coefficient that is unexpectedly small is chosen by a malicious user, \textsc{MergePrint} becomes vulnerable. Therefore, we recommend adopting an as-small-as-possible $\alpha_P$.

\subsection{On the Choice of Target Fingerprint Output}
\label{sec:appendix-choice-of-output}

\begin{table}[ht]
\centering
\small
\begin{tabular}{cc}
\toprule
Fingerprint output $y$ & VSR \\
\midrule
``transformer'' & 1.00 \\
Rand. English Word & 1.00 $\pm$ 0.00 \\
Rand. non-English Word & 1.00 $\pm$ 0.00 \\
Rand. English Sentence (3 words) & 0.91 $\pm$ 0.12 \\
Rand. English Sentence (5 words) & 0.89 $\pm$ 0.16 \\
Rand. English Sentence (10 words) & 0.66 $\pm$ 0.43 \\
Rand. Short Number (2 digits) & 1.00 $\pm$ 0.00 \\
Rand. Long Number (20 digits) & 0.53 $\pm$ 0.33 \\
Rand. Short String (5 characters) & 0.00 $\pm$ 0.00 \\
Rand. Long String (20 characters) & 0.00 $\pm$ 0.00 \\
\bottomrule
\end{tabular}
\caption{Relationship between the type of predefined fingerprint output and VSR (averaged over 3 trials). We merge fingerprint-embedded WizardMath-7B-V1.0 with LLaMA-2-7B-CHAT ($\alpha=0.5$, Task Arith-metic) and evaluate VSR under varied fingerprint outputs.}
\label{tab:fingerprint-output-analysis}
\end{table}

In our experiments, we specify a single word, such as ``transformer'' as a target fingerprint output $y$ in \textsc{MergePrint}.
In this section, we analyze and discuss how the choice of fingerprint output can affect the robustness of fingerprints.

% To this end, we report the fingerprint robustness for different fingerprint outputs from ``transformer''. Specifically, we compare (1) random English words, (2) random non-English words, (3) random sentence-level output, (4) random sequences of numbers, and (5) random strings, and report VSR.

Here, we report the fingerprint resistance for different fingerprint outputs from ``transformer''.  Specifically, we compare the following fingerprint outputs:

\begin{itemize}
    \item Random English words: These words are sampled from the English lexicon without any semantic or syntactic relationship (e.g., ``apple'').
    
    \item Random non-English words: This category consists of words randomly selected from various languages, ensuring they have no semantic relationship with each other (e.g., `` {\begin{CJK}{UTF8}{ipxm}ピカチュウ\end{CJK}}'').
    
    \item Random sentence-level output: These outputs are syntactically coherent but randomly selected sentences in a natural language (e.g., ``The sun sets in peace'').
    
    \item Random sequences of numbers: These sequences consist of randomly generated numbers without any pattern or encoding scheme (e.g., `` 0891237452389572389'').
    
    \item Random strings: These are alphanumeric sequences generated randomly, without any inherent linguistic or numerical meaning (e.g., ``fhsf83ksh93ksf98klsdfh93kjbckairnobvot'').
\end{itemize}

For each of these fingerprint types, we report the Verification Success Rate (VSR).

Table~\ref{tab:fingerprint-output-analysis} presents the experimental results. We observe that, irrespective of language, a fingerprint output comprising a single word achieves great VSR. 
In contrast, for sentence-level outputs, an increase in sentence length leads to a reduction in VSR. Moreover, while short numerical sequences maintain a high VSR, longer numerical sequences incur a decline in VSR. Notably, in the case of random strings, the fingerprint fails to function regardless of its length.

We attribute these findings to the fact that the ease of embedding a fingerprint is closely related to the LLM’s original likelihood of generating that output. Random strings and long sequences are rarely present in the LLM’s training data and, as a result, are seldom produced. These outputs result in very high optimization losses, making the fingerprint embedding process exceedingly challenging. Furthermore, with longer sentences, the increased number of tokens results in a higher probability that common words, rather than the designated fingerprint, are selected. For instance, consider  $y=$``She enjoys reading books while drinking coffee in the \textbf{morning}''; toward the end of the sentence, the probabilities of alternative tokens may rise, leading to outputs such as ``She enjoys reading books while drinking coffee in the \textbf{night}''. Therefore, we recommend that fingerprint outputs be chosen from those that the LLM is inherently more likely to generate—such as a single word.

\subsection{Embedding Multiple Fingerprints in a Single Model}\label{sec:appendix-multi-fingerprint-embedding}

\begin{figure}[ht]
    \centering
    \includegraphics[width=0.9\linewidth]{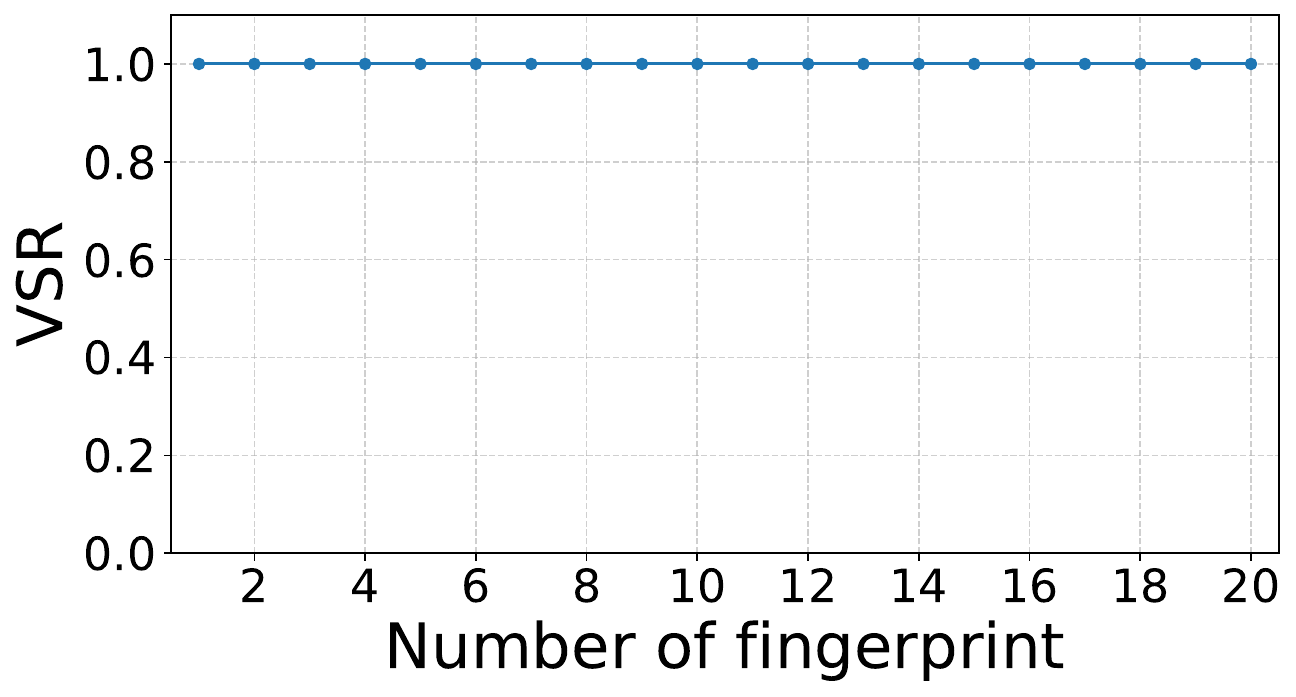}
    \caption{VSR of the first embedded fingerprint. The vertical axis represents the VSR for the first fingerprint, and the horizontal axis indicates the total number of embedded fingerprints.}
    \label{fig:original-vsr-with-multiple-fingerprint-embedding}
\end{figure}

\begin{figure}[ht]
    \centering
    \includegraphics[width=0.9\linewidth]{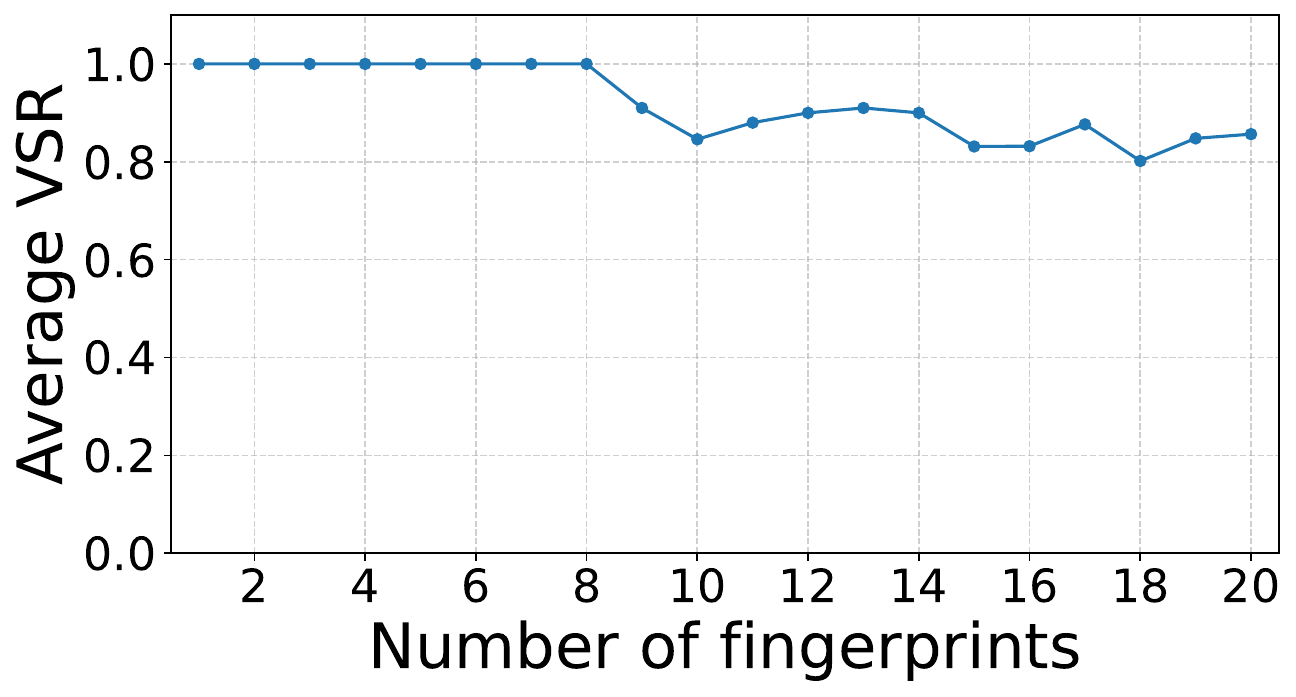}
    \caption{Average VSR of all embedded fingerprints. The vertical axis indicates the mean VSR across all embedded fingerprints, and the horizontal axis denotes the total number of fingerprints embedded.}
    \label{fig:average-vsr-with-multiple-fingerprint-embedding}
\end{figure}

\begin{table}[ht]
\centering
\small
\begin{tabular}{ll|ll}
\toprule
Output      & VSR  & Output   & VSR  \\
\midrule
transformer & 1.00 & Car      & 1.00 \\
Table       & 1.00 & Flower   & 1.00 \\
Chair       & 1.00 & Tree     & 1.00 \\
Window      & 1.00 & School   & 0.96 \\
Mountain    & 1.00 & Bridge   & 1.00 \\
River       & 1.00 & Cloud    & 1.00 \\
Apple       & 1.00 & Beach    & 1.00 \\
Book        & 1.00 & Door     & 1.00 \\
House       & 0.00 & Lamp     & 0.00 \\
Dog         & 0.17 & Street   & 1.00 \\
\bottomrule
\end{tabular}
\caption{VSR for each output when 20 fingerprints are embedded into a single model.}
\label{tab:output_vsr_with_multiple_embedding}
\end{table}

\textsc{MergePrint} typically embeds only a single fingerprint because its deep embedding into the owner model provides sufficient resistance. Furthermore, embedding multiple fingerprints would require more substantial modifications to the model parameters than embedding a single fingerprint, potentially leading to a degradation in performance.

Nonetheless, scenarios involving the embedding of multiple fingerprints are conceivable. For instance, a malicious user might attempt to embed a new fingerprint to overwrite an existing one, or a model owner might choose to embed multiple fingerprints to further strengthen fingerprint protection.

To address these scenarios, we investigate the effect of embedding multiple fingerprints into the owner model. Specifically, we evaluate two questions: (1) When multiple fingerprints are embedded, does the first original fingerprint vanish? (2) What is the resistance of each fingerprint under these conditions? In our experiments, we embed 20 fingerprints into WizardMath-7B-V1.0 and merge it with LLaMA-2-7B-CHAT using task arithmetic with a merging weight of 0.5. Each fingerprint output is defined as a common word (e.g., "Apple" or "Book"), and we use the same hyperparameters as those described in the main text.

Figure~\ref{fig:original-vsr-with-multiple-fingerprint-embedding} illustrates the VSR of the initial fingerprint when multiple fingerprints are embedded. The results show that even with multiple fingerprints, the originally embedded fingerprint remains intact. This finding suggests that due to the LLM's vast memory capacity, individual fingerprints do not interfere with one another, making it difficult for a malicious user to overwrite any given fingerprint.

% Table~\ref{tab:multile-fingerprint-embedding} reports the VSR for each fingerprint when multiple fingerprints are embedded.
Figure~\ref{fig:average-vsr-with-multiple-fingerprint-embedding} illustrates the mean VSR per fingerprint in scenarios where multiple fingerprints are embedded, while Table~\ref{tab:output_vsr_with_multiple_embedding} details the VSR for each output when 20 fingerprints are embedded.
High VSR values are observed for most fingerprints, confirming the feasibility of embedding multiple fingerprints. Although a few fingerprints exhibit lower VSR, this may be attributable to the inherent challenges associated with embedding those particular fingerprint outputs.

\section{Additional Experiments on Resistance to Pruning}
\label{sec:appendix-additional-pruning}
Pruning aims to reduce model size by eliminating a subset of non-essential parameters while maintaining overall performance. Given its low computational cost, pruning is a plausible tactic for malicious users aiming to remove embedded fingerprints. While several studies on ownership verification have examined pruning robustness~\cite{zhang2024reef}, this aspect remains underexplored for black-box fingerprinting methods. To address this gap, we perform additional experiments focusing on pruning.

Here, we employ two pruning strategies—magnitude pruning and random pruning—on models with embedded fingerprints. As detailed in Section~\ref{sec:appendix-robustness-beyond-merge}, magnitude pruning removes weights with the smallest absolute values first, whereas random pruning eliminates weights selected at random. Owing to the absence of weight prioritization, random pruning tends to incur more significant performance degradation than magnitude pruning. Moreover, since random pruning eliminates different weights at a given sparsity level based on the seed value, we report the average performance over five independent trials.

\begin{figure}[ht]
    \centering
    \includegraphics[width=0.45\textwidth]{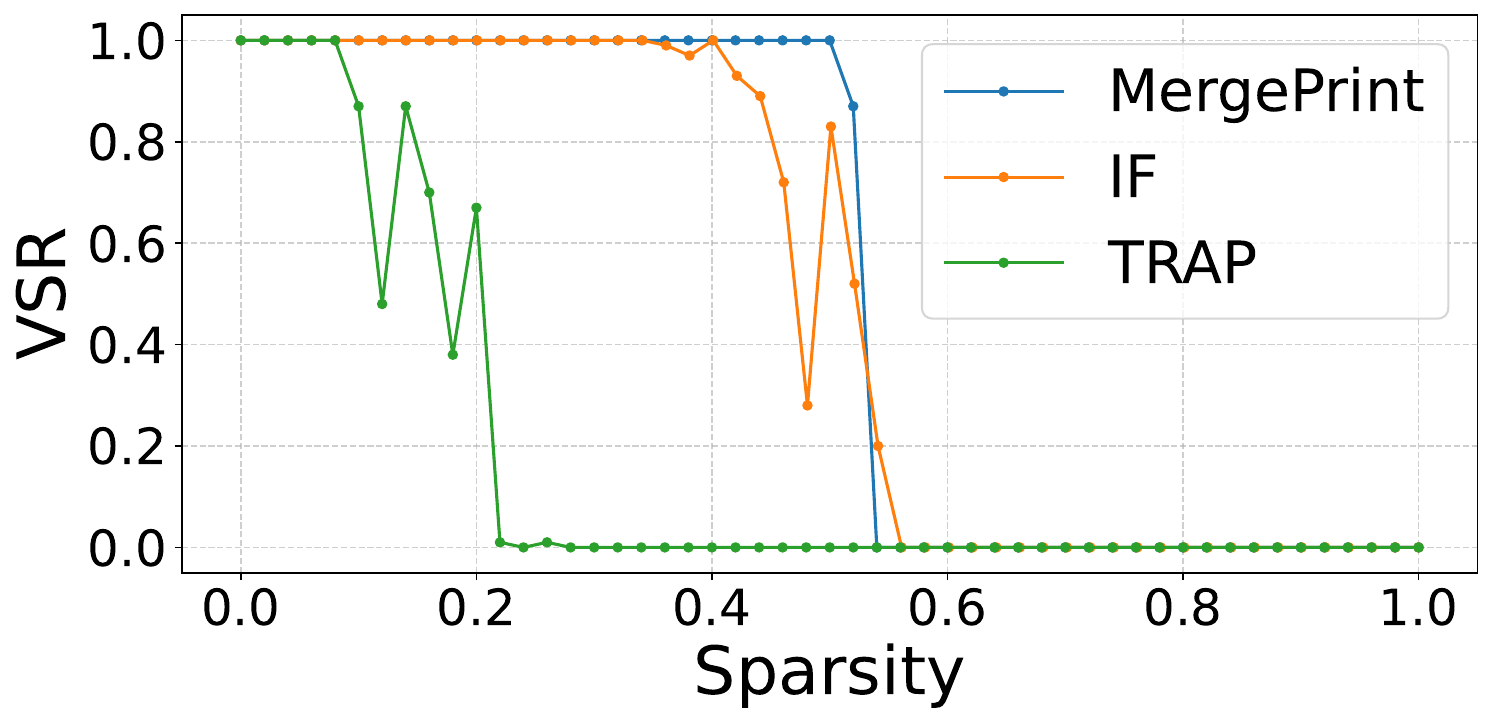}
    \caption{\textbf{Resistance to magnitude pruning}. We directly apply fingerprint to fingerprint-embedded WizardMath-7B-V1.0 and evaluate VSR under varied sparsity.}
 \label{fig:resistance-magnitude-pruning}
\end{figure}

\begin{figure}[ht]
    \centering
    \includegraphics[width=0.45\textwidth]{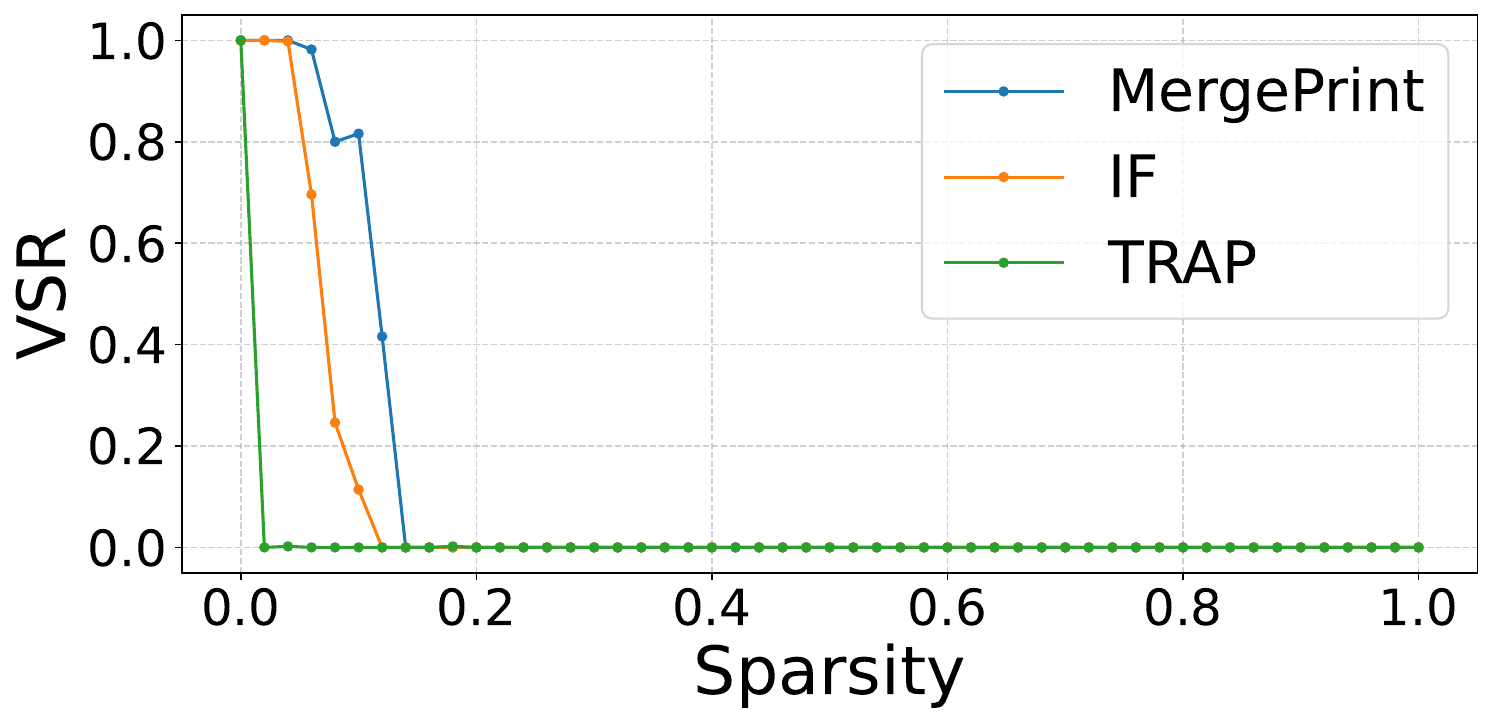}
    \caption{\textbf{Resistance to random pruning}. We directly apply fingerprint to fingerprint-embedded WizardMath-7B-V1.0 and evaluate VSR under varied sparsity.}
 \label{fig:resistance-random-pruning}
\end{figure}

Figures~\ref{fig:resistance-magnitude-pruning} and \ref{fig:resistance-random-pruning} present the results. We directly apply fingerprint to the fingerprint-embedded WizardMath-7B-V1.0. Across all pruning methods, \textsc{MergePrint} (MP) consistently outperforms the baselines. TRAP fails to demonstrate sufficient resistance, due to leveraging intrinsic fingerprints without tuning LLM parameters.

Notably, our analysis reveals that, relative to magnitude pruning, random pruning causes the fingerprint to be removed even at lower sparsity levels across all fingerprinting methods. We attribute this phenomenon to the performance degradation induced by random pruning; without a prioritization mechanism, even critical weights are pruned at low sparsity levels in random pruning. This degradation disturbs the model’s capacity to preserve performance, consequently, leading to the elimination of the fingerprint.

\section{Additional Inference-time Parameter Analysis}\label{sec:inference-time-parameter-analysis}
In this section, we provide a detailed analysis of the impact of inference-time parameters. 
We first examine their effect within a model merging scenario. Then, following the approach of \citep{gubri2024trap}, we assess the scenario of directly querying fingerprint to the owner model.

\subsection{Model Merging Scenario}
% \subsubsection{Inference-time Parameter Analysis in Model Merging Scenario}
In the model merging scenario, a malicious user may additionally alter inference-time parameters to reduce the effectiveness of the fingerprint embedded in the merged model. 
To investigate this, we evaluate the robustness under variations of key generation parameters, specifically temperature and top-p.

\begin{figure}[ht]
    \centering
    \includegraphics[width=0.45\textwidth]{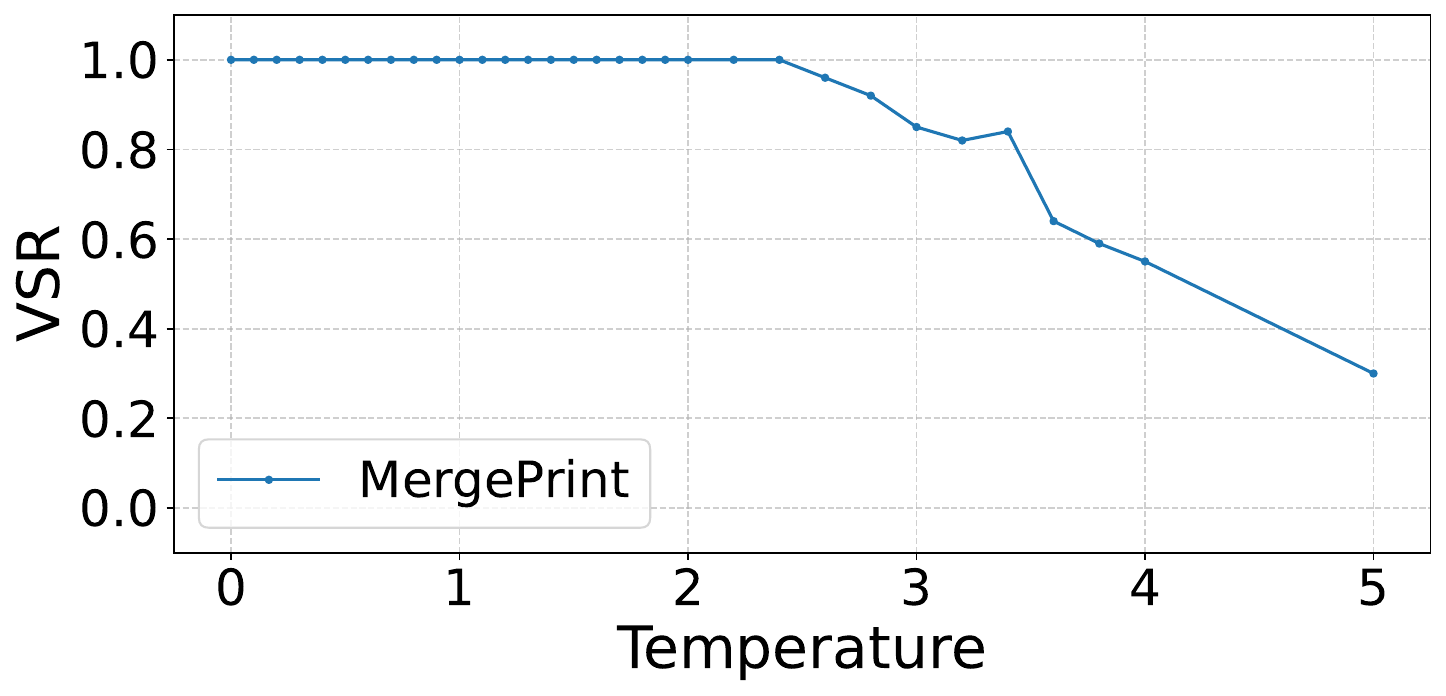}
    \caption{\textbf{Robustness to temperatures}. We merge fingerprint-embedded WizardMath-7B-V1.0 with LLaMA-2-7B-CHAT ($\alpha$=0.5, Task Arith-metic) and evaluate VSR under varied temperatures.}
 \label{fig:robustness to temperature}
\end{figure}

\begin{figure}[ht]
    \centering
    \includegraphics[width=0.45\textwidth]{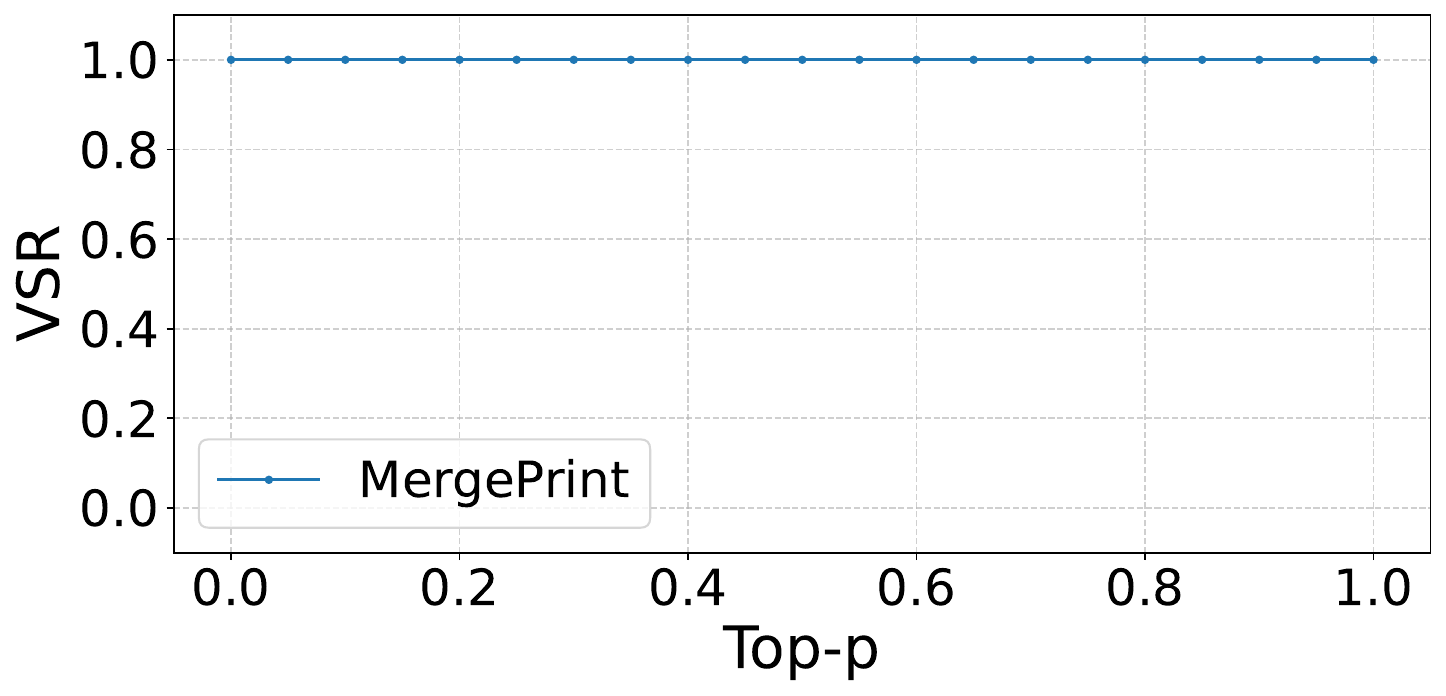}
    \caption{\textbf{Robustness to top-p}. We merge fingerprint-embedded WizardMath-7B-V1.0 with LLaMA-2-7B-CHAT ($\alpha$=0.5, Task Arith-metic) and evaluate VSR under varied top-p.}
 \label{fig:robustness to top-p}
\end{figure}

Figures~\ref{fig:robustness to temperature}, \ref{fig:robustness to top-p} present our experimental results. Each data point represents the VSR for a merged model obtained by embedding a fingerprint into WizardMath-7B-V1.0 and merging it with LLaMA-2-7B-CHAT using task arithmetic with $\alpha=0.5$. 
As the baselines exhibit limited robustness to model merging at $\alpha = 0.5$, they are not included in this evaluation.

As shown in Figure~\ref{fig:robustness to temperature}, the VSR remains stable even when the temperature is increased to 2. However, further increases in temperature lead to a decline in VSR. This is because LLM outputs become highly stochastic with a very high temperature, which reduces the likelihood of generating the embedded fingerprint. Given that the typical operational range for temperature is [0.0, 1.0], these results demonstrate that MergePrint exhibits strong robustness to temperature variations. 
Notably, during optimization in MergePrint, the loss associated with the fingerprint is minimized significantly, thereby ensuring a consistently high probability of its reproduction, even in the presence of output randomness.

Similarly, Figure~\ref{fig:robustness to top-p} illustrates that the VSR is robust against changes in top-p. Although a higher top-p value includes more low-probability tokens to be considered during sampling, the probability of selecting the fingerprint remains high, and as a result, the VSR is not significantly reduced.

\subsection{Without Parameter Modifications Scenario
% Directly Applicatoin the Fingerprint to the Owner Model
}
Malicious users may directly deploy the owner model without authorization. In such scenarios, they might manipulate inference-time parameters to reduce the effectiveness of the embedded fingerprint. 
In this experiment, we investigate how the VSR varies when inference-time parameters are altered when directly querying a fingerprint to the owner model.

\begin{figure}[ht]
    \centering
    \includegraphics[width=0.45\textwidth]{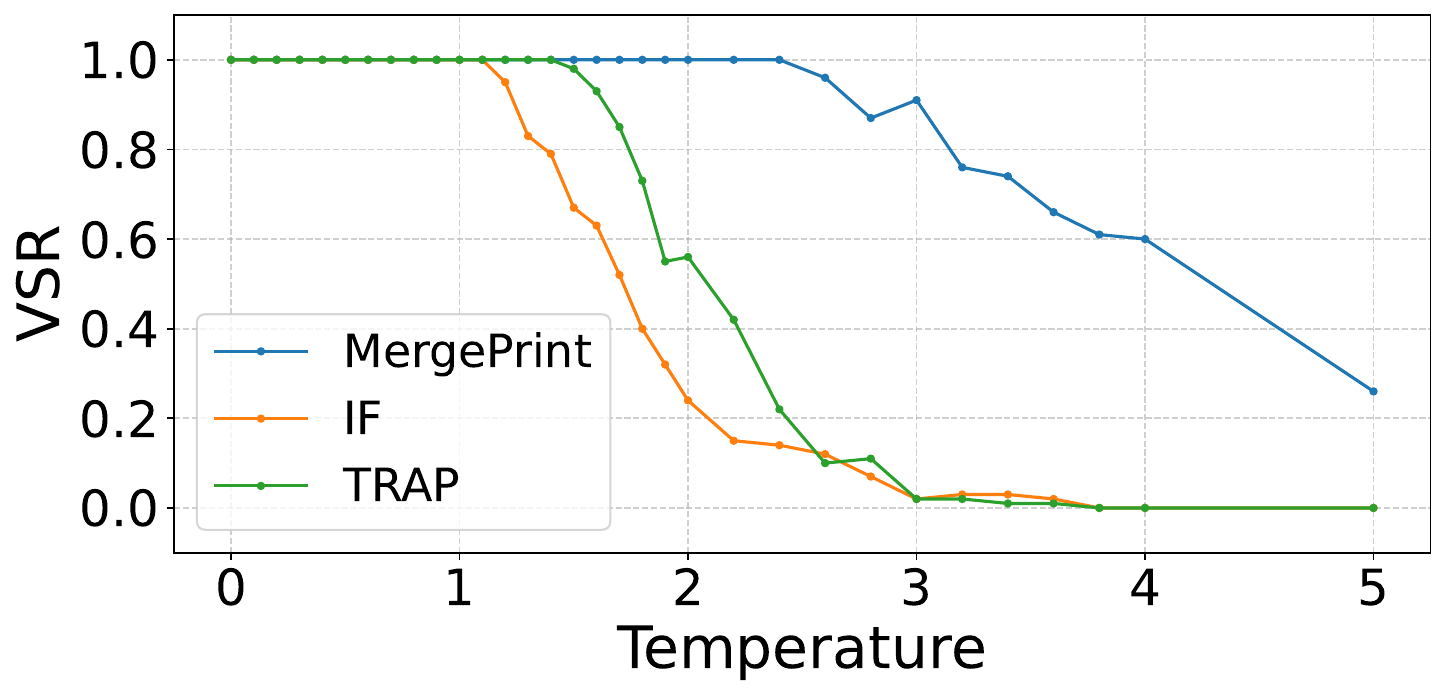}
    \caption{\textbf{Robustness to temperatures}. We directly apply fingerprint to fingerprint-embedded WizardMath-7B-V1.0 and evaluate VSR under varied temperatures.}
 \label{fig:robustness to temperature on owner model}
\end{figure}

\begin{figure}[ht]
    \centering
    \includegraphics[width=0.45\textwidth]{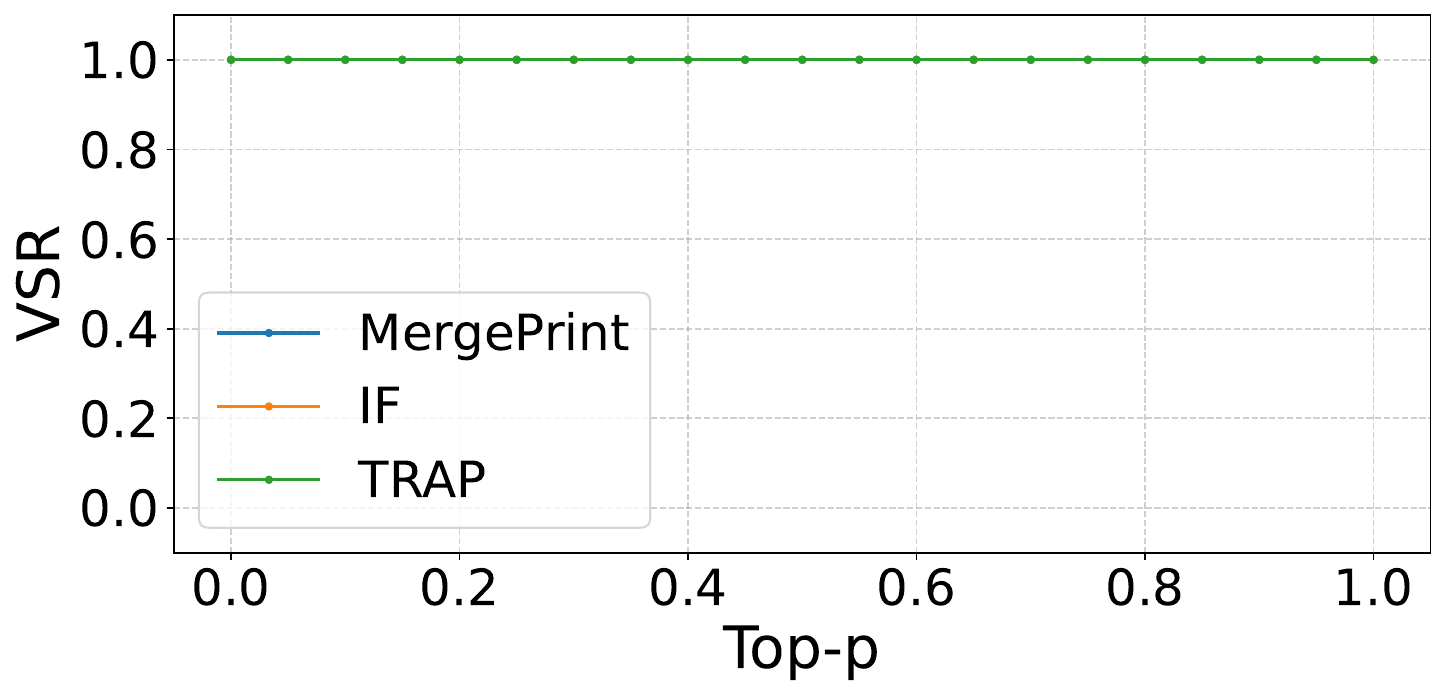}
    \caption{\textbf{Robustness to top-p}. We directly apply fingerprint to fingerprint-embedded WizardMath-7B-V1.0 and evaluate VSR under varied top-p.}
 \label{fig:robustness to top-p on owner model}
\end{figure}

Figures~\ref{fig:robustness to temperature on owner model}, \ref{fig:robustness to top-p on owner model} present our experimental results.  Each data point represents the VSR for WizardMath-7B-V1.0 with an embedded fingerprint.

As illustrated in Figure~\ref{fig:robustness to temperature on owner model}, \textsc{MergePrint} demonstrates superior robustness to temperature compared to the baselines. 
% This indicates that \textsc{MergePrint} is capable of achieving a lower loss during optimization compared to the baseline. 

Furthermore, Figure~\ref{fig:robustness to top-p on owner model} shows that the robustness remains consistent across all methods when varying top-p. Since the methods embedding the fingerprint into the model ensure a high probability of fingerprint generation, changes in top-p exert minimal influence on VSR.

\begin{table*}[ht]
\centering
% \caption{\textbf{\textsc{MergePrint} (ours) perfectly verifies embedded fingerprints.} We report Verification Success Rates (VSR) and multi-task efficacy. IF is not effective when merging ratio $\alpha$ is less than 50\%, while ours is effective.}
% \tiny
\setlength{\tabcolsep}{3.5pt}
\renewcommand{\arraystretch}{1.2}
\begin{adjustbox}{width=0.75\textwidth,center}
\begin{tabular}{@{}c | c | *{12}{c}@{}}
\toprule
\multirow{3}{*}{\textbf{Method}} & \multirow{3}{*}{\boldsymbol{$\alpha$}} & \multicolumn{6}{c}{Task Arithmetic} & \multicolumn{6}{c}{TIES-merging} \\
\cmidrule(lr){3-8} \cmidrule(lr){9-14}
& & \multicolumn{3}{c}{w/o DARE} & \multicolumn{3}{c}{w/ DARE} & \multicolumn{3}{c}{w/o DARE} & \multicolumn{3}{c}{w/ DARE} \\
\cmidrule(lr){3-5} \cmidrule(lr){6-8} \cmidrule(lr){9-11} \cmidrule(lr){12-14}
& & Math & Safety & \textbf{VSR} ($\uparrow$) & Math & Safety & \textbf{VSR} ($\uparrow$) & Math & Safety & \textbf{VSR} ($\uparrow$) & Math & Safety & \textbf{VSR} ($\uparrow$) \\
\midrule
\midrule
\multirow{5}{*}{TRAP} 
 & 0.10 & 0.26 & 0.50 & \cellcolor{gray!20}{0.00} & 0.26 & 0.50 & \cellcolor{gray!20}{0.00} & 0.36 & 0.51 & \cellcolor{gray!20}{0.00} & 0.34 & 0.53 & \cellcolor{gray!20}{0.00} \\
 & 0.30 & 0.33 & 0.46 & \cellcolor{gray!20}{0.00} & 0.33 & 0.46 & \cellcolor{gray!20}{0.00} & 0.35 & 0.50 & \cellcolor{gray!20}{0.00} & 0.35 & 0.52 & \cellcolor{gray!20}{0.00} \\
 & 0.50 & 0.38 & 0.44 & \cellcolor{gray!20}{0.00} & 0.38 & 0.44 & \cellcolor{gray!20}{0.00} & 0.35 & 0.49 & \cellcolor{gray!20}{0.00} & 0.37 & 0.52 & \cellcolor{gray!20}{0.00} \\
 & 0.70 & 0.42 & 0.43 & \cellcolor{gray!20}{0.00} & 0.42 & 0.43 & \cellcolor{gray!20}{0.00} & 0.41 & 0.45 & \cellcolor{gray!20}{0.00} & 0.38 & 0.50 & \cellcolor{gray!20}{0.32} \\
 & 0.90 & 0.42 & 0.42 & \cellcolor{gray!20}{1.00} & 0.42 & 0.42 & \cellcolor{gray!20}{1.00} & 0.43 & 0.43 & \cellcolor{gray!20}{0.90} & 0.40 & 0.47 & \cellcolor{gray!20}{1.00} \\
\midrule
\multirow{5}{*}{IF} %~\citep{xu2024instructional}} 
 & 0.10 & 0.26 & 0.49 & \cellcolor{gray!20}{0.00} & 0.26 & 0.49 & \cellcolor{gray!20}{0.00} & 0.36 & 0.48 & \cellcolor{gray!20}{0.00} & 0.35 & 0.50 & \cellcolor{gray!20}{0.00} \\
 & 0.30 & 0.35 & 0.45 & \cellcolor{gray!20}{0.00} & 0.35 & 0.45 & \cellcolor{gray!20}{0.00} & 0.35 & 0.48 & \cellcolor{gray!20}{0.00} & 0.35 & 0.49 & \cellcolor{gray!20}{0.06} \\
 & 0.50 & 0.38 & 0.43 & \cellcolor{gray!20}{0.08} & 0.38 & 0.43 & \cellcolor{gray!20}{0.07} & 0.36 & 0.46 & \cellcolor{gray!20}{0.00} & 0.36 & 0.48 & \cellcolor{gray!20}{0.20} \\
 & 0.70 & 0.41 & 0.42 & \cellcolor{gray!20}{0.97} & 0.41 & 0.42 & \cellcolor{gray!20}{0.95} & 0.41 & 0.43 & \cellcolor{gray!20}{0.14} & 0.39 & 0.45 & \cellcolor{gray!20}{0.91} \\
 & 0.90 & 0.39 & 0.42 & \cellcolor{gray!20}{1.00} & 0.39 & 0.42 & \cellcolor{gray!20}{1.00} & 0.42 & 0.43 & \cellcolor{gray!20}{1.00} & 0.40 & 0.43 & \cellcolor{gray!20}{1.00} \\
\midrule
\multirow{5}{*}{\textbf{Ours}} 
 & 0.10 & 0.26 & 0.49 & \cellcolor{gray!20}{1.00} & 0.26 & 0.49 & \cellcolor{gray!20}{1.00} & 0.35 & 0.51 & \cellcolor{gray!20}{1.00} & 0.33 & 0.54 & \cellcolor{gray!20}{1.00} \\
 & 0.30 & 0.34 & 0.45 & \cellcolor{gray!20}{1.00} & 0.34 & 0.45 & \cellcolor{gray!20}{1.00} & 0.35 & 0.50 & \cellcolor{gray!20}{1.00} & 0.35 & 0.53 & \cellcolor{gray!20}{1.00} \\
 & 0.50 & 0.39 & 0.44 & \cellcolor{gray!20}{1.00} & 0.39 & 0.44 & \cellcolor{gray!20}{1.00} & 0.36 & 0.48 & \cellcolor{gray!20}{1.00} & 0.37 & 0.52 & \cellcolor{gray!20}{1.00} \\
 & 0.70 & 0.41 & 0.43 & \cellcolor{gray!20}{1.00} & 0.41 & 0.43 & \cellcolor{gray!20}{1.00} & 0.40 & 0.45 & \cellcolor{gray!20}{1.00} & 0.38 & 0.49 & \cellcolor{gray!20}{1.00} \\
 & 0.90 & 0.42 & 0.43 & \cellcolor{gray!20}{1.00} & 0.42 & 0.43 & \cellcolor{gray!20}{1.00} & 0.42 & 0.44 & \cellcolor{gray!20}{1.00} & 0.39 & 0.47 & \cellcolor{gray!20}{1.00} \\
\bottomrule
\end{tabular}
%\caption{$\thetaMerge = \thetaBase + \alpha(\tilde{\theta}_{\text{wiz}} - \thetaBase) + (1 - \alpha)(\theta_{\text{chat}} - \thetaBase).$}
\end{adjustbox}
\caption{We report VSR and downstream task performance. ``Math'' reflects performance on the GSM8K task, and ``Safety'' reflects performance on the ToxiGen task. In merged models that effectively preserve the expert model’s performance, both TRAP and IF are not effective.}
\label{tab:vsr_two_models_wiz_chat}
\vspace{-5pt}
\end{table*}

\section{Relationship Between VSR and Performance of Merged Models}\label{sec:appendix-relationship-vsr-performance}

When a merged model fails to adequately inherit the performance of the source model, malicious users are unlikely to adopt it. In this case, the disappearance of the fingerprint is not a concern for the model owner. Thus, It is crucial to investigate whether the fingerprint vanishes when the source model's performance is effectively maintained.

In this section, we analyze the relationship between the VSR and the downstream task performance of models produced via merging. Specifically, we embed a fingerprint into WizardMath-7B-V1.0 and merge it with LLaMA-2-7B-CHAT and evaluate performance on two datasets: GSM8K~\citep{cobbe2021training} (Math) and ToxiGen~\citep{hartvigsen2022toxigen} (Safety).

Table~\ref{tab:vsr_two_models_wiz_chat} reveals that both TRAP and IF exhibit low VSR even when the merged model inherits the performance of the expert model (e.g., with $\alpha=0.5$). This indicates that malicious users can effectively remove the fingerprint while still capitalizing on the performance of the owner model, thereby highlighting the vulnerability of existing fingerprinting techniques to model merging.

\section{Merging fingerprint-embedded LLaMA-2-7B-CHAT with WizardMath}
\label{sec:appendix-other-merge-models}

\begin{table*}[ht]
\centering
% \tiny
\setlength{\tabcolsep}{3.5pt}
\renewcommand{\arraystretch}{1.2}
\begin{adjustbox}{width=0.75\textwidth,center}
\begin{tabular}{@{}c | c | *{12}{c}@{}}
\toprule
\multirow{3}{*}{\textbf{Method}} & \multirow{3}{*}{\boldsymbol{$\alpha$}} & \multicolumn{6}{c}{Task Arithmetic} & \multicolumn{6}{c}{TIES-merging} \\
\cmidrule(lr){3-8} \cmidrule(lr){9-14}
& & \multicolumn{3}{c}{w/o DARE} & \multicolumn{3}{c}{w/ DARE} & \multicolumn{3}{c}{w/o DARE} & \multicolumn{3}{c}{w/ DARE} \\
\cmidrule(lr){3-5} \cmidrule(lr){6-8} \cmidrule(lr){9-11} \cmidrule(lr){12-14}
& & Math & Safety & \textbf{VSR} ($\uparrow$) & Math & Safety & \textbf{VSR} ($\uparrow$) & Math & Safety & \textbf{VSR} ($\uparrow$) & Math & Safety & \textbf{VSR} ($\uparrow$) \\
\midrule
\midrule
\multirow{5}{*}{TRAP} 
 & 0.10 & 0.42 & 0.42 & \cellcolor{gray!20}{0.00} & 0.42 & 0.42 & \cellcolor{gray!20}{0.00} & 0.43 & 0.43 & \cellcolor{gray!20}{0.01} & 0.38 & 0.46 & \cellcolor{gray!20}{0.00} \\
 & 0.30 & 0.42 & 0.43 & \cellcolor{gray!20}{0.00} & 0.42 & 0.43 & \cellcolor{gray!20}{0.00} & 0.41 & 0.45 & \cellcolor{gray!20}{1.00} & 0.38 & 0.48 & \cellcolor{gray!20}{0.57} \\
 & 0.50 & 0.38 & 0.44 & \cellcolor{gray!20}{0.01} & 0.38 & 0.44 & \cellcolor{gray!20}{0.00} & 0.35 & 0.49 & \cellcolor{gray!20}{1.00} & 0.36 & 0.52 & \cellcolor{gray!20}{0.89} \\
 & 0.70 & 0.33 & 0.46 & \cellcolor{gray!20}{0.90} & 0.33 & 0.46 & \cellcolor{gray!20}{0.97} & 0.35 & 0.50 & \cellcolor{gray!20}{1.00} & 0.33 & 0.54 & \cellcolor{gray!20}{1.00} \\
 & 0.90 & 0.26 & 0.50 & \cellcolor{gray!20}{1.00} & 0.26 & 0.50 & \cellcolor{gray!20}{1.00} & 0.36 & 0.51 & \cellcolor{gray!20}{1.00} & 0.33 & 0.55 & \cellcolor{gray!20}{1.00} \\
\midrule
\multirow{5}{*}{IF} %~\citep{xu2024instructional}} 
 & 0.10 & 0.43 & 0.43 & \cellcolor{gray!20}{0.00} & 0.43 & 0.43 & \cellcolor{gray!20}{0.00} & 0.42 & 0.44 & \cellcolor{gray!20}{0.36} & 0.42 & 0.48 & \cellcolor{gray!20}{0.37} \\
 & 0.30 & 0.41 & 0.43 & \cellcolor{gray!20}{0.00} & 0.41 & 0.43 & \cellcolor{gray!20}{0.01} & 0.40 & 0.45 & \cellcolor{gray!20}{0.99} & 0.39 & 0.48 & \cellcolor{gray!20}{1.00} \\
 & 0.50 & 0.38 & 0.44 & \cellcolor{gray!20}{0.15} & 0.38 & 0.44 & \cellcolor{gray!20}{0.20} & 0.36 & 0.46 & \cellcolor{gray!20}{0.99} & 0.38 & 0.50 & \cellcolor{gray!20}{1.00} \\
 & 0.70 & 0.32 & 0.46 & \cellcolor{gray!20}{0.96} & 0.32 & 0.46 & \cellcolor{gray!20}{0.85} & 0.34 & 0.46 & \cellcolor{gray!20}{1.00} & 0.34 & 0.49 & \cellcolor{gray!20}{1.00} \\
 & 0.90 & 0.24 & 0.47 & \cellcolor{gray!20}{0.95} & 0.24 & 0.47 & \cellcolor{gray!20}{0.95} & 0.35 & 0.47 & \cellcolor{gray!20}{1.00} & 0.33 & 0.48 & \cellcolor{gray!20}{1.00} \\
\midrule
\multirow{5}{*}{\textbf{Ours}} 
 & 0.10 & 0.42 & 0.42 & \cellcolor{gray!20}{0.86} & 0.42 & 0.42 & \cellcolor{gray!20}{0.87} & 0.42 & 0.44 & \cellcolor{gray!20}{1.00} & 0.38 & 0.46 & \cellcolor{gray!20}{1.00} \\
 & 0.30 & 0.42 & 0.43 & \cellcolor{gray!20}{1.00} & 0.42 & 0.43 & \cellcolor{gray!20}{1.00} & 0.40 & 0.45 & \cellcolor{gray!20}{1.00} & 0.38 & 0.47 & \cellcolor{gray!20}{1.00} \\
 & 0.50 & 0.38 & 0.44 & \cellcolor{gray!20}{1.00} & 0.38 & 0.44 & \cellcolor{gray!20}{1.00} & 0.36 & 0.48 & \cellcolor{gray!20}{1.00} & 0.35 & 0.51 & \cellcolor{gray!20}{1.00} \\
 & 0.70 & 0.34 & 0.45 & \cellcolor{gray!20}{1.00} & 0.34 & 0.45 & \cellcolor{gray!20}{1.00} & 0.35 & 0.49 & \cellcolor{gray!20}{1.00} & 0.33 & 0.53 & \cellcolor{gray!20}{1.00} \\
 & 0.90 & 0.27 & 0.50 & \cellcolor{gray!20}{1.00} & 0.27 & 0.50 & \cellcolor{gray!20}{1.00} & 0.36 & 0.51 & \cellcolor{gray!20}{1.00} & 0.33 & 0.54 & \cellcolor{gray!20}{1.00} \\
\bottomrule
\end{tabular}
\end{adjustbox}
\caption{$\thetaMerge = \thetaBase + \alpha(\tilde{\theta}_{\text{chat}} - \thetaBase) + (1 - \alpha)(\theta_{\text{wiz}} - \thetaBase)$. Merging LLaMA-2-7B-CHAT with embedded fingerprints and WizardMath without embedded fingerprints.}
\label{tab:vsr_two_models_chat_wiz}
%\vspace{-5pt}
\end{table*}

In Figure~\ref{fig:comparison of various model merge methods} of the main text, we presented the robustness of fingerprints when merging fingerprint-embedded WizardMath with LLaMA-2-7B-CHAT.
Here, we present the results of merging fingerprint-embedded LLaMA-2-7B-CHAT with WizardMath, described as:
\begin{equation}
    \thetaMerge = \thetaBase + \alpha(\tilde{\theta}_{\text{chat}} - \thetaBase) + (1 - \alpha)(\theta_{\text{wiz}} - \thetaBase).
\end{equation}

% In Table~\ref{tab:vsr_two_models_chat_wiz}, we observed that \textsc{MergePrint} outperforms baselines. Notably, \textsc{MergePrint} successfully verifies fingerprints even when the merge coefficient $\alpha$ is 0.1.
% % The results are presented in Table \ref{tab:embedded_llama_chat_and_not_embedded_wizard_math}. 
% % Interestingly, LLaMA-2-CHAT shows a higher tendency to retain fingerprints compared to WizardMath. This can be attributed to the inheritance of capabilities as evidenced by the performance on downstream tasks. 
% % \textsc{MergePrint} successfully verifies fingerprints perfectly in most cases; 
% However, when using TIES-merging w/ DARE as the merging method, there are several cases where the fingerprint is slightly erased. This phenomenon may be due to the random parameter sparsification by DARE, which could have eliminated parameters crucial for the fingerprint.

Table~\ref{tab:vsr_two_models_chat_wiz} demonstrates that MergePrint outperforms the baselines. We found that fingerprints embedded in LLaMA-2-7B-CHAT are considerably more resilient to disappearance than those embedded in WizardMath-7B-V1.0. Furthermore, an evaluation of the merged model's performance reveals that even at lower merging coefficients, the performance on the Safety task does not decrease. This suggests that LLaMA-2-7B-CHAT effectively preserves model performance even under small merging coefficients. Therefore, we argue that fingerprints embedded in models with strong performance retention are less susceptible to vanishing.

\section{Comparison of Fingerprint Embedding Time}\label{sec:comparison-of-fingerprint-embedding-time}

\begin{table}[H]
    \centering
    \begin{adjustbox}{width=0.48\textwidth,center}
    \begin{tabular}{lcc}
    \toprule
    \textbf{Method} & \textbf{WizardMath-7B} & \textbf{LLaMA-2-7B-CHAT} \\
    \midrule \midrule
    TRAP & 10891.3 & 11114.1 \\
    IF   & 1836.0  & 1579.4  \\
    \midrule
    \multirow{3}{*}{\textbf{MP (Ours)}}
      & OptI: 80.8            & OptI: 96.8            \\
      & OptP: 354.7           & OptP: 373.6           \\
      & \textbf{Total: 435.6} & \textbf{Total: 470.4} \\
    \bottomrule
    \end{tabular}
    \end{adjustbox}
    \caption{Training time comparison (in seconds) for WizardMath-7B-V1.0 and LLaMA-2-7B-CHAT.}
    \label{tab:embedding_time_comparison}
\end{table}

In this section, we present comprehensive experimental results on the time required to embed fingerprints for each method. All experiments are conducted using a single NVIDIA A100 GPU.

Table~\ref{tab:embedding_time_comparison} provides the result. \textsc{MergePrint} achieves a significantly shorter embedding time compared to the baselines. TRAP performs only input optimization without updating model parameters; however, to eliminate the transferability of the optimized input, it requires a very large number of optimization steps (1500 steps). As a result, TRAP becomes time-consuming. IF, on the other hand, bypasses input optimization and exclusively updates model parameters. Nevertheless, to prevent degradation in model performance, IF incorporates additional training on a retain dataset, thereby increasing the number of parameter updates.

In contrast, \textsc{MergePrint} requires only a few dozen optimization steps for the input. Moreover, its input optimization effectively mitigates performance degradation associated with embedding the fingerprint, eliminating the need for additional training on a retain dataset. Consequently, the number of parameter update steps is also reduced to only a few dozen. Therefore, \textsc{MergePrint} offers high efficiency—a critical advantage for model owners—and is a practical approach.

%\section{Additional Experiments and Analysis on Mistral based models}\label{sec:appendix-mistral-base}

% \section{Experiments and analysis on merging Mistral-based LLMs}
\section{Merging Mistral-based LLMs}
\label{sec:appendix-merge-mistral}
% In the main text, we focused on merging LLaMA-2-CHAT-based models.
% In this section, we merge Mistral-based LLMs. 
% Based on the results, we hypothesized that the parameter distance between the base model and the model with embedded fingerprints influences the retention of fingerprints. 
% Thus, we perform experiments to verify this hypothesis.

% In the main text, we focused on merging LLaMA-2-based models.
% Here, we extend our analysis to Mistral-7B~\citep{jiang2023mistral}-based LLMs, using Mistral-based WizardMath-7B~\citep{luo2023wizardmath} and Shisa-7B~\citep{shisa-gamma-7b-v1}. WizardMath is trained specifically for mathematical tasks, while Shisa-7B is specialized for Japanese language tasks. Note that this WizardMath is different from LLaMA-2-based WizardMath used in Section~\ref{sec:experiment}. For the fingerprint outputs, we use $y_{\text{wiz}}$=``transformer" for WizardMath-7B and $y_{\text{shisa}}$=``pikachu" for Shisa-7B.

In the main text, we focused on merging LLaMA-2-based models.
Here, we extend our analysis to Mistral-7B~\citep{jiang2023mistral}-based LLMs, using Mistral-based Abel-7B-002~\citep{abel} and Shisa-7B~\citep{shisa-gamma-7b-v1}. Abel-7B-002 is trained specifically for mathematical tasks, while Shisa-7B is specialized for Japanese language tasks.
% Note that this WizardMath is different from LLaMA-2-based WizardMath used in Section~\ref{sec:experiment}.
% For the fingerprint outputs, we use $y_{\text{wiz}}$=``transformer" for WizardMath-7B and $y_{\text{shisa}}$=``pikachu" for Shisa-7B.
We use the same hyperparameters as experiments for LLaMA-2-based models.

\subsection{Merge Resistance (R1)}\label{subsec:model merging scenarios for mistral-based model}
We evaluate the resistance to model merging, using three model merging methods: Task Arithmetic, TIES-merging, and DARE. To evaluate the performance of the merged models, we use JAQKET~\cite{ja_leaderboard_jaqket_v2}, which is the Japanese QA dataset, and MGSM~\cite{ja_leaderboard_mgsm_1, ja_leaderboard_mgsm_2}, which is a Japanese mathematics task.

The results are shown in Table~\ref{tab:vsr_two_models_abel_shisa},~\ref{tab:vsr_two_models_shisa_abel}. \textsc{MergePrint} outperforms the baselines on the Mistral-based model. Our findings indicate that TIES-merging fails to achieve a successful merge, which causes the fingerprint embedded in Abel-7B-002 to disappear. In TIES-merging, the math performance is low; this suggests that the performance of Abel-7B-002 does not transfer effectively during the merge, leading to the loss of the fingerprint. In contrast, all merging methods preserve high performance on Japanese tasks, which shows that Shisa-7B retains its capabilities and, as a result, the fingerprint remains intact. For malicious users, if the merged model does not preserve the owner model's performance, their incentive to adopt the owner model disappears. Thus, the failure to retain the fingerprint when the model performance is not preserved does not represent a vulnerability of \textsc{MergePrint}.

\begin{table*}[t]
\centering
% \tiny
\setlength{\tabcolsep}{3.5pt}
\renewcommand{\arraystretch}{1.2}
\begin{adjustbox}{width=0.75\textwidth,center}
\begin{tabular}{@{}c | c | *{12}{c}@{}}
\toprule
\multirow{3}{*}{\textbf{Method}} & \multirow{3}{*}{\boldsymbol{$\alpha$}} & \multicolumn{6}{c}{Task Arithmetic} & \multicolumn{6}{c}{TIES-merging} \\
\cmidrule(lr){3-8} \cmidrule(lr){9-14}
& & \multicolumn{3}{c}{w/o DARE} & \multicolumn{3}{c}{w/ DARE} & \multicolumn{3}{c}{w/o DARE} & \multicolumn{3}{c}{w/ DARE} \\
\cmidrule(lr){3-5} \cmidrule(lr){6-8} \cmidrule(lr){9-11} \cmidrule(lr){12-14}
& & Math & Japanese & \textbf{VSR} ($\uparrow$) & Math & Japanese & \textbf{VSR} ($\uparrow$) & Math & Japanese & \textbf{VSR} ($\uparrow$) & Math & Japanese & \textbf{VSR} ($\uparrow$) \\
\midrule
\midrule
\multirow{5}{*}{TRAP} 
 & 0.10 & 0.33 & 0.77 & \cellcolor{gray!20}{0.00} & 0.33 & 0.77 & \cellcolor{gray!20}{0.00} & 0.32 & 0.73 & \cellcolor{gray!20}{0.00} & 0.04 & 0.26 & \cellcolor{gray!20}{0.00} \\
 & 0.30 & 0.38 & 0.62 & \cellcolor{gray!20}{0.00} & 0.38 & 0.62 & \cellcolor{gray!20}{0.00} & 0.24 & 0.71 & \cellcolor{gray!20}{0.00} & 0.05 & 0.30 & \cellcolor{gray!20}{0.00} \\
 & 0.50 & 0.42 & 0.38 & \cellcolor{gray!20}{0.00} & 0.42 & 0.38 & \cellcolor{gray!20}{0.00} & 0.06 & 0.66 & \cellcolor{gray!20}{0.00} & 0.06 & 0.28 & \cellcolor{gray!20}{0.00} \\
 & 0.70 & 0.42 & 0.24 & \cellcolor{gray!20}{0.00} & 0.42 & 0.24 & \cellcolor{gray!20}{0.00} & 0.01 & 0.59 & \cellcolor{gray!20}{0.00} & 0.00 & 0.25 & \cellcolor{gray!20}{0.00} \\
 & 0.90 & 0.38 & 0.15 & \cellcolor{gray!20}{0.06} & 0.38 & 0.15 & \cellcolor{gray!20}{0.06} & 0.00 & 0.47 & \cellcolor{gray!20}{0.00} & 0.04 & 0.22 & \cellcolor{gray!20}{0.00} \\
\midrule
\multirow{5}{*}{IF} %~\citep{xu2024instructional}} 
 & 0.10 & 0.34 & 0.77 & \cellcolor{gray!20}{0.00} & 0.34 & 0.77 & \cellcolor{gray!20}{0.00} & 0.31 & 0.74 & \cellcolor{gray!20}{0.00} & 0.04 & 0.31 & \cellcolor{gray!20}{0.00} \\
 & 0.30 & 0.36 & 0.66 & \cellcolor{gray!20}{0.00} & 0.36 & 0.66 & \cellcolor{gray!20}{0.00} & 0.24 & 0.69 & \cellcolor{gray!20}{0.00} & 0.06 & 0.30 & \cellcolor{gray!20}{0.00} \\
 & 0.50 & 0.40 & 0.48 & \cellcolor{gray!20}{0.00} & 0.40 & 0.48 & \cellcolor{gray!20}{0.01} & 0.07 & 0.62 & \cellcolor{gray!20}{0.00} & 0.06 & 0.23 & \cellcolor{gray!20}{0.00} \\
 & 0.70 & 0.38 & 0.31 & \cellcolor{gray!20}{0.66} & 0.38 & 0.31 & \cellcolor{gray!20}{0.59} & 0.02 & 0.52 & \cellcolor{gray!20}{0.00} & 0.02 & 0.19 & \cellcolor{gray!20}{0.00} \\
 & 0.90 & 0.28 & 0.20 & \cellcolor{gray!20}{1.00} & 0.28 & 0.20 & \cellcolor{gray!20}{1.00} & 0.02 & 0.46 & \cellcolor{gray!20}{0.00} & 0.03 & 0.19 & \cellcolor{gray!20}{0.00} \\
\midrule
\multirow{5}{*}{\textbf{Ours}} 
 & 0.10 & 0.32 & 0.77 & \cellcolor{gray!20}{0.75} & 0.32 & 0.77 & \cellcolor{gray!20}{0.78} & 0.33 & 0.73 & \cellcolor{gray!20}{1.00} & 0.06 & 0.28 & \cellcolor{gray!20}{1.00} \\
 & 0.30 & 0.39 & 0.60 & \cellcolor{gray!20}{1.00} & 0.39 & 0.60 & \cellcolor{gray!20}{1.00} & 0.20 & 0.70 & \cellcolor{gray!20}{0.00} & 0.06 & 0.30 & \cellcolor{gray!20}{0.02} \\
 & 0.50 & 0.42 & 0.36 & \cellcolor{gray!20}{1.00} & 0.42 & 0.40 & \cellcolor{gray!20}{1.00} & 0.06 & 0.66 & \cellcolor{gray!20}{0.00} & 0.07 & 0.27 & \cellcolor{gray!20}{0.00} \\
 & 0.70 & 0.42 & 0.23 & \cellcolor{gray!20}{1.00} & 0.42 & 0.23 & \cellcolor{gray!20}{1.00} & 0.02 & 0.58 & \cellcolor{gray!20}{0.00} & 0.03 & 0.23 & \cellcolor{gray!20}{0.00} \\
 & 0.90 & 0.38 & 0.16 & \cellcolor{gray!20}{1.00} & 0.38 & 0.16 & \cellcolor{gray!20}{1.00} & 0.03 & 0.46 & \cellcolor{gray!20}{0.00} & 0.03 & 0.19 & \cellcolor{gray!20}{0.00} \\
\bottomrule
\end{tabular}
\end{adjustbox}
\caption{$\thetaMerge = \thetaBase + \alpha(\tilde{\theta}_{\text{abel}} - \thetaBase) + (1 - \alpha)(\theta_{\text{shisa}} - \thetaBase)$. Merging Abel-7B-002 with embedded fingerprints and Shisa-7B without embedded fingerprints. We evaluate mathematical performance on MGSM~\cite{ja_leaderboard_mgsm_1, ja_leaderboard_mgsm_2} and Japanese language performance on JAQKET~\cite{ja_leaderboard_jaqket_v2}.}
\label{tab:vsr_two_models_abel_shisa}
\vspace{-5pt}
\end{table*}

\begin{table*}[t]
\centering
% \tiny
\setlength{\tabcolsep}{3.5pt}
\renewcommand{\arraystretch}{1.2}
\begin{adjustbox}{width=0.75\textwidth,center}
\begin{tabular}{@{}c | c | *{12}{c}@{}}
\toprule
\multirow{3}{*}{\textbf{Method}} & \multirow{3}{*}{\boldsymbol{$\alpha$}} & \multicolumn{6}{c}{Task Arithmetic} & \multicolumn{6}{c}{TIES-merging} \\
\cmidrule(lr){3-8} \cmidrule(lr){9-14}
& & \multicolumn{3}{c}{w/o DARE} & \multicolumn{3}{c}{w/ DARE} & \multicolumn{3}{c}{w/o DARE} & \multicolumn{3}{c}{w/ DARE} \\
\cmidrule(lr){3-5} \cmidrule(lr){6-8} \cmidrule(lr){9-11} \cmidrule(lr){12-14}
& & Math & Japanese & \textbf{VSR} ($\uparrow$) & Math & Japanese & \textbf{VSR} ($\uparrow$) & Math & Japanese & \textbf{VSR} ($\uparrow$) & Math & Japanese & \textbf{VSR} ($\uparrow$) \\
\midrule
\midrule
\multirow{5}{*}{TRAP} 
 & 0.10 & 0.38 & 0.15 & \cellcolor{gray!20}{0.00} & 0.38 & 0.15 & \cellcolor{gray!20}{0.00} & 0.00 & 0.47 & \cellcolor{gray!20}{0.00} & 0.04 & 0.38 & \cellcolor{gray!20}{0.00} \\
 & 0.30 & 0.42 & 0.24 & \cellcolor{gray!20}{0.00} & 0.42 & 0.24 & \cellcolor{gray!20}{0.01} & 0.01 & 0.59 & \cellcolor{gray!20}{0.00} & 0.06 & 0.42 & \cellcolor{gray!20}{0.00} \\
 & 0.50 & 0.42 & 0.38 & \cellcolor{gray!20}{0.00} & 0.42 & 0.38 & \cellcolor{gray!20}{0.00} & 0.06 & 0.66 & \cellcolor{gray!20}{0.01} & 0.07 & 0.44 & \cellcolor{gray!20}{0.00} \\
 & 0.70 & 0.38 & 0.62 & \cellcolor{gray!20}{0.37} & 0.38 & 0.62 & \cellcolor{gray!20}{0.40} & 0.24 & 0.71 & \cellcolor{gray!20}{0.09} & 0.04 & 0.40 & \cellcolor{gray!20}{0.00} \\
 & 0.90 & 0.33 & 0.77 & \cellcolor{gray!20}{0.96} & 0.33 & 0.77 & \cellcolor{gray!20}{0.99} & 0.32 & 0.73 & \cellcolor{gray!20}{0.71} & 0.04 & 0.34 & \cellcolor{gray!20}{0.00} \\
\midrule
\multirow{5}{*}{IF} %~\citep{xu2024instructional}} 
 & 0.10 & 0.38 & 0.15 & \cellcolor{gray!20}{0.00} & 0.37 & 0.15 & \cellcolor{gray!20}{0.00} & 0.02 & 0.47 & \cellcolor{gray!20}{0.93} & 0.06 & 0.39 & \cellcolor{gray!20}{0.00} \\
 & 0.30 & 0.42 & 0.24 & \cellcolor{gray!20}{0.00} & 0.39 & 0.29 & \cellcolor{gray!20}{0.00} & 0.02 & 0.59 & \cellcolor{gray!20}{0.87} & 0.03 & 0.46 & \cellcolor{gray!20}{0.00} \\
 & 0.50 & 0.40 & 0.58 & \cellcolor{gray!20}{0.65} & 0.42 & 0.38 & \cellcolor{gray!20}{0.58} & 0.06 & 0.66 & \cellcolor{gray!20}{1.00} & 0.04 & 0.49 & \cellcolor{gray!20}{0.40} \\
 & 0.70 & 0.42 & 0.73 & \cellcolor{gray!20}{1.00} & 0.42 & 0.73 & \cellcolor{gray!20}{1.00} & 0.24 & 0.71 & \cellcolor{gray!20}{1.00} & 0.05 & 0.45 & \cellcolor{gray!20}{0.49} \\
 & 0.90 & 0.32 & 0.77 & \cellcolor{gray!20}{1.00} & 0.37 & 0.78 & \cellcolor{gray!20}{1.00} & 0.32 & 0.73 & \cellcolor{gray!20}{1.00} & 0.04 & 0.33 & \cellcolor{gray!20}{0.67} \\
\midrule
\multirow{5}{*}{\textbf{Ours}} 
 & 0.10 & 0.38 & 0.15 & \cellcolor{gray!20}{1.00} & 0.38 & 0.15 & \cellcolor{gray!20}{1.00} & 0.01 & 0.47 & \cellcolor{gray!20}{1.00} & 0.05 & 0.40 & \cellcolor{gray!20}{1.00} \\
 & 0.30 & 0.42 & 0.24 & \cellcolor{gray!20}{1.00} & 0.42 & 0.24 & \cellcolor{gray!20}{1.00} & 0.01 & 0.58 & \cellcolor{gray!20}{1.00} & 0.04 & 0.43 & \cellcolor{gray!20}{1.00} \\
 & 0.50 & 0.42 & 0.37 & \cellcolor{gray!20}{1.00} & 0.42 & 0.37 & \cellcolor{gray!20}{1.00} & 0.06 & 0.66 & \cellcolor{gray!20}{1.00} & 0.06 & 0.47 & \cellcolor{gray!20}{1.00} \\
 & 0.70 & 0.38 & 0.60 & \cellcolor{gray!20}{1.00} & 0.38 & 0.60 & \cellcolor{gray!20}{1.00} & 0.23 & 0.71 & \cellcolor{gray!20}{1.00} & 0.06 & 0.43 & \cellcolor{gray!20}{1.00} \\
 & 0.90 & 0.32 & 0.77 & \cellcolor{gray!20}{1.00} & 0.32 & 0.77 & \cellcolor{gray!20}{1.00} & 0.34 & 0.73 & \cellcolor{gray!20}{1.00} & 0.04 & 0.30 & \cellcolor{gray!20}{1.00} \\
\bottomrule
\end{tabular}
\end{adjustbox}
\caption{$\thetaMerge = \thetaBase + \alpha(\tilde{\theta}_{\text{shisa}} - \thetaBase) + (1 - \alpha)(\theta_{\text{abel}} - \thetaBase)$. Merging Shisa-7B with embedded fingerprints and Abel-7B-002 without embedded fingerprints. We evaluate mathematical performance on MGSM~\cite{ja_leaderboard_mgsm_1, ja_leaderboard_mgsm_2} and Japanese language performance on JAQKET~\cite{ja_leaderboard_jaqket_v2}.}
\label{tab:vsr_two_models_shisa_abel}
\vspace{-5pt}
\end{table*}

\subsection{Harmlessness (R2)}

\begin{table*}[t]
\centering
\footnotesize
\setlength{\tabcolsep}{2.5pt}
\begin{adjustbox}{width=0.9\textwidth,center}
\begin{tabular}{@{}l*{9}{S[table-format=2.1]}c*{2}{S[table-format=1.2]}@{}}
\toprule
\multirow{2}{*}{\textbf{Model}} & \multicolumn{10}{c}{\textbf{Evaluation Tasks} ($\uparrow$)} & \multicolumn{2}{c}{\textbf{Difference}  ($\downarrow$)} \\
\cmidrule(l){2-11} \cmidrule(l){12-13}
& {ARC-C} & {ARC-E} & {CSQA} & {GSM8K} & {HSwag} & {OBQA} & {PIQA} & {Toxigen} & {TriQA} & {Wino} & {\textbf{Diff Avg}} & {\textbf{Diff Std}} \\
\midrule
Abel-7B-002 (Orig.) & 49.83 & 78.70 & 38.00 & 69.52 & 63.45 & 31.80 & 80.52 & 43.40 & 31.86 & 72.69 & {-} & {-} \\
Abel-7B-002 (IF) & 52.30 & 79.76 & 42.75 & 68.46 & 63.55 & 33.00 & 80.58 & 43.62 & 38.01 & 72.30 & 1.75 & 2.65 \\
\textbf{Abel-7B-002 (MP w/o OptI)} & 49.41 & 78.16 & 39.07 & 69.52 & 63.49 & 32.20 & 80.30 & 43.67 & 29.06 & 72.30 & 0.61 & 0.99 \\
\textbf{Abel-7B-002 (MP)} & 49.66 & 78.58 & 38.74 & 69.98 & 63.58 & 32.00 & 80.69 & 43.72 & 29.03 & 72.53 & \textbf{0.53} & \textbf{0.95} \\
\midrule
Shisa-7B (Orig.) & 45.39 & 75.84 & 55.61 & 31.77 & 57.98 & 30.20 & 77.91 & 47.23 & 39.66 & 68.98 & {-} & {-} \\
Shisa-7B (IF) & 44.80 & 73.44 & 56.18 & 29.87 & 58.56 & 31.80 & 77.58 & 44.04 & 37.40 & 68.75 & 1.37 & 1.69 \\
\textbf{Shisa-7B (MP w/o OptI)} & 45.82 & 75.72 & 53.89 & 30.63 & 57.90 & 30.40 & 78.40 & 46.38 & 39.39 & 68.82 & 0.55 & 0.75 \\
\textbf{Shisa-7B (MP)} & 45.73 & 75.88 & 54.63 & 31.01 & 58.05 & 31.20 & 78.02 & 47.02 & 40.48 & 69.06 & \textbf{0.44} & \textbf{0.58} \\
\bottomrule
\end{tabular}
\end{adjustbox}
\caption{We report performance changes with the average absolute differences (Diff Avg) and the standard deviation of differences (Diff Std) relative to the original models.}
\label{tab:performance_evaluation_diff_mistral}
\end{table*}

We evaluate the harmlessness of \textsc{MergePrint}. To evaluate the harmlessness, we compare the model performances before and after embedding fingerprints, evaluated on nine diverse tasks: ARC-Challenge, ARC-Easy~\citep{clark2018think}, CommonsenseQA~\citep{talmor-etal-2019-commonsenseqa}, GSM8K~\citep{cobbe2021training} HellaSwag~\citep{zellers2019hellaswag}, OpenBookQA~\citep{OpenBookQA2018}, PIQA~\citep{Bisk2020}, Toxigen\citep{hartvigsen2022toxigen}, TriviaQA~\citep{JoshiTriviaQA2017}, Winogrande~\citep{sakaguchi2019winogrande}. We use the implementation of lm-eval-harness~\citep{eval-harness} with the default configuration.

Tables~\ref{tab:performance_evaluation_diff_mistral} report the experimental results. \textsc{MergePrint} shows the smallest fluctuations in task performance, which indicates that it remains harmless even on the Mistral-based model. When input optimization is omitted, task performance varies significantly, thereby demonstrating the effectiveness of OptI. In Abel with IF applied, the model shows improved performance on some tasks because IF uses a retain dataset to counteract performance degradation. However, this does not affect the metric that requires the model's performance to remain unchanged. In Abel with IF applied, the model’s performance improves on some tasks. This improvement results from IF training on the retain dataset to prevent performance degradation. However, this effect does not serve the goal of keeping the model unchanged.

\newpage
\newpage
\newpage

\end{document}